\documentclass{iopjournal}

\usepackage{graphicx}%
\usepackage{multirow}%
\usepackage{amsmath,amssymb,amsfonts}%
\usepackage{amsthm}%
\usepackage{mathrsfs}%
\usepackage[title]{appendix}%
\usepackage{xcolor}%
\usepackage{textcomp}%
\usepackage{manyfoot}%
\usepackage{booktabs}%
\usepackage{algorithm}%
\usepackage{algorithmicx}%
\usepackage{algpseudocode}%
\usepackage{listings}%

\usepackage{afterpage}
\usepackage[
    defaultcolor=red,
]{changes}
\definechangesauthor[name=X, color=blue]{+}

\begin{document}
\articletype{Paper} 

\title{Quantum-safe IPsec in the banking industry}


\author{
Rafael~J.~Vicente$^{1,*}$\orcid{0000-0001-6334-639X},
Jaime~G\'omez~Garc\'ia$^2$,
Juan~P.~Brito$^{1,3}$\orcid{0000-0002-0970-6638},
Yorlandy~Lobaina$^1$,
Jaime~S.~Buruaga$^1$\orcid{0000-0002-9502-9978},
Daniel~G\'omez~Aguado$^2$\orcid{0000-0002-8093-6065},
Miguel~\'Angel~S\'anchez~Serrano$^4$,
Sim\'on~Ovsyannikov$^4$,
Salah~Gherdaoui$^4$,
Jean-S\'ebastien~Pegon$^5$,
Marco~Cofano$^6$ and
Vicente~Mart\'in$^{1,3}$\orcid{0000-0002-2559-3979}
}

\affil{$^1$Center~for~Computational~Simulation,~Universidad~Polit\'ecnica~de~Madrid,~Madrid,~Spain}
\affil{$^2$Santander~Digital~Services,~Madrid,~Spain}
\affil{$^3$QoolNet~Sociedad~Limitada,~Madrid,~Spain}
\affil{$^4$Cisco,~San~Jose,~California,~USA}
\affil{$^5$ID~Quantique,~Geneva,~Switzerland}
\affil{$^6$Luxquanta,~Barcelona,~Spain}
\affil{$^*$Author~to~whom~any~correspondence~should~be~addressed.}

\email{rafaelj.vicente@upm.es}


\begin{abstract}
The emergence of Cryptographically Relevant Quantum Computers (CRQCs) presents a critical threat to classical cryptographic systems, particularly widely adopted protocols such as RSA, Diffie–Hellman (DH), and Elliptic Curve Cryptography (ECC). Given their extensive use in the financial sector, the advent of quantum adversaries compels banking institutions to proactively develop and adopt quantum-safe communication mechanisms. This paper introduces a hybrid quantum-safe architecture, orchestrated via Software-Defined Networking (SDN) key distribution. The proposed framework enables the early integration of Classical Cryptography (CC), Quantum Key Distribution (QKD), and Post-Quantum Cryptography (PQC) within a Dynamic Multipoint Virtual Private Network (DMVPN) environment, providing highly scalable, full-mesh, site-to-site encrypted communications for enterprise networks. This is particularly relevant at a time when PQC algorithms have not yet been incorporated into finalized IPsec standards. The architecture has been validated across a five-node testbed comprising three physical nodes within a campus network in Madrid and two private-cloud nodes located in the north of Spain and Mexico. The deployment leverages a heterogeneous mix of physical and virtual devices, diverse technology providers, Discrete Variable QKD (DV-QKD) and Continuous Variable QKD (CV-QKD) implementations, and mutually incompatible key-delivery interfaces (ETSI004, ETSI014 and Cisco SKIP), demonstrating flexibility, scalability, and interoperability across environments. Through this framework, we demonstrate that quantum-safe communication in financial networks is not only technically feasible but also scalable, interoperable, and resilient. The proposed architecture establishes a robust, flexible, and future-proof foundation for secure financial communications in the era of quantum computing, ensuring multivendor, multi-interface, and multitechnology interoperability while minimizing disruption to existing infrastructures and workflows.
\end{abstract}

\section{Introduction}\label{sec1}

Information security relies on cryptographic techniques that ensure the confidentiality, authenticity, integrity, and non-repudiation of data transmitted over the internet. Every transaction performed, every message exchanged, every financial operation depends on cryptographic protocols to guarantee trust between peers in digital processes. These mechanisms are essential for creating a secure communication infrastructure, particularly important in critical sectors such as banking.

The arrival of quantum computing introduces a new threat to current cryptographic systems. CRQCs are expected to be developed in the 2030s~\cite{globalrisk}. These computers have the ability to break current public-key cryptographic algorithms such as Rivest, Shamir, Adleman (RSA)~\cite{RSA}, Diffie-Hellman~(DH)~\cite{DH}, and Elliptic Curve Diffie-Hellman~(ECDH)~\cite{MillerECC}~\cite{Koblitz1987}, exposing the systems that use them to potentially severe risks. As a result, strategic sectors such as finance must adopt quantum-safe solutions to ensure the integrity and confidentiality of their communications. 
The banking sector has been actively seeking new strategies to safeguard communications to anticipate the emergence of potentially harmful threats. The shift to quantum-safe cryptography is not just a precautionary measure, but an essential step in preserving the security of financial systems. The two most widely adopted approaches are Post-Quantum Cryptography (PQC), which is based on mathematical algorithms that, to date, resist all known classical and quantum attacks; and Quantum Key Distribution (QKD), capable of performing symmetric key distillation using algorithms based on the laws of quantum mechanics, capable of guaranteeing so-called  information-theoretic security (ITS) for the key-distribution stage, under the standard assumptions of QKD security proofs.

PQC algorithms, including lattice-based, hash-based, and multivariate schemes, are based on mathematical problems without known quantum or classical algorithms able to solve them efficiently. Some implementations of those schemes are currently under standardization by the U.S. National Institute of Standards and Technology (NIST). The first ones that have been recently approved are Module-Lattice-Based Key-Encapsulation Mechanism (ML-KEM)~\cite{FIPS203} for key encapsulation mechanism and Module-Lattice-Based Digital Signature Algorithm (ML-DSA)~\cite{FIPS204} and Stateless Hash-Based Digital Signature Algorithm (SLH-DSA)~\cite{FIPS205} for digital signature. These algorithms are designed to replace current public-key systems in a way that remains secure against both classical and quantum adversaries. In parallel, QKD enables two modules to exchange symmetric keys in an ITS manner, which represents the maximum level of protection in a security environment. Several financial institutions have begun piloting QKD-based technologies.

Moreover, QKD allows two modules to establish symmetric keys using principles of quantum mechanics and can provide security from an information-theoretic perspective for the process of key distribution between two non-adjacent trusted nodes.

This work presents a hybrid solution that combines classical cryptography (CC), QKD, and PQC, enabling the adoption of secure quantum technologies in banking environments. This solution has been tested on a real financial network composed of five trusted nodes, including three nodes implemented with physical systems and two nodes deployed on virtualized systems in private cloud locations. This hybrid topology offers a realistic scenario for evaluating these new technologies in production systems while reflecting the heterogeneity of the financial infrastructure.

A distinct feature of this approach is the use of Software-Defined Networking (SDN)~\cite{SDN_arch_ONF, Kreutz_SDN_Survey, Engineering_SDN_QKD} for dynamic management of quantum-safe technologies.  SDN decouples the control and data planes, offering programmable management of network behavior. This facilitates the integration of different cryptographic policies, rapid reconfiguration, and the combined use of classical and quantum security mechanisms without the need to modify legacy systems. The proposed  architecture follows a trusted‑node paradigm for QKD. It implies that intermediate nodes are assumed to be physically secured and operated under appropriate procedural controls. Key Management Systems (KMS) are trusted to provide secure key storage and distribution, while the SDN controller is trusted to orchestrate key allocation and network resources correctly. Initial authentication of classical channels relies on conventional pre‑shared keys mechanisms.

The solution's design prioritizes compatibility with different systems that follow QKD and PQC standards, thus minimizing the changes required to adopt these technologies in the existing infrastructure. The proposed solution has been integrated with the current banking network infrastructure without affecting its workflows. The result is a cryptoagile system that supports the deployment of hybrid cryptography schemes while preserving interoperability with classic systems and avoiding vendor lock-in. 

Financial networks often use IPsec VPNs for intersite communication, relying on the Internet Key Exchange (IKEv2) protocol and Diffie–Hellman (DH) or Elliptic Curve DH (ECDH) key exchange to establish secure tunnels. However, these components are vulnerable to quantum attacks. Previous work has investigated the replacement of DH with either QKD or PQC~\cite{alia2024100}, but to the best of our knowledge, it was done for isolated links and no solution has been tested on a scalable setup, able to provide full mesh connectivity among all network nodes of a large corporation as we have done.

Our implementation extends this model by integrating Dynamic Multipoint VPN (DMVPN) with quantum-safe key exchange mechanisms. Each tunnel is protected using at least one form of quantum-resistant cryptography: QKD for metro networks and PQC for long-haul protection with cloud nodes. The architecture supports multivendor deployments and incorporates both Discrete Variable (DV-QKD) and Continuous Variable QKD (CV-QKD) devices, increasing the options available for infrastructure design while maintaining robust cryptographic diversity and vendor independence.

Furthermore, cryptoagility is achieved through intelligent SDN management, allowing the system to switch between cryptographic modes based on link capability, risks, and performance constraints such as Secret Key Rate (SKR). This approach guarantees uninterrupted security for future deployments while accommodating the ongoing evolution of cryptographic standards.

To assess the security properties of the proposed architecture, we explicitly delineate adversarial capabilities and trust assumptions. We consider a external adversary able to observe, intercept, store, replay, and actively manipulate classical communications transiting untrusted wide‑area networks, including IKEv2/IPsec exchanges and control‑plane traffic carried over shared operator infrastructures. The attacker is further assumed to possess long‑term storage (enabling harvest‑now–decrypt‑later strategies) and, eventually, quantum computational capabilities. Under this model, the system is designed to prevent derivation of tunnel keys and to preserve session confidentiality using quantum‑safe mechanisms. This proposal aims to mitigate passive eavesdropping, man‑in‑the‑middle attacks during key establishment, and future quantum attacks against classical public‑key schemes, even when the adversary controls the surrounding transport fabric.

We assume a hardened, trusted perimeter for the Trusted Nodes, their associated internal components including KMS functions, SDN controller and optical transporte elements. Management and data interfaces are authenticated, physical access to the devices is controlled, and operational configurations are correct and monitored. Consequently, insider threats, physical compromise of trusted components, and malicious behavior within the KMS or SDN control plane are out of scope and expected to be mitigated with complementary safeguards, consistent with prior SDN–QKD architectures \cite{b.mendezQuantumResistantSoftware2026}. This boundary preserves security guarantees while enabling operator‑grade orchestration at scale.

In sort, this solution proposes a scalable and standards based model for securing banking networks against quantum threats. By combining classical cryptography with PQC and QKD technologies through an SDN framework and aligning ourselves with current operational practices, we demonstrate that the transition to secure quantum communication is achievable. These results offer a model for financial institutions willing to prepare their security infrastructures for the future, looking ahead to the arrival of CRQCs.

\section{Background}

IPsec is a widely adopted suite of protocols designed to secure communications at the IP layer. It is most commonly used to establish Virtual Private Networks (VPNs), either between two network gateways or between a remote client and an enterprise gateway. IPsec relies on the Internet Key Exchange version 2 protocol (IKEv2)~\cite{rfc5996} to negotiate cryptographic parameters and establish secure communication channels. IKEv2 separates the initial negotiation into two exchanges: IKE\_SA\_INIT establishes a secure, authenticated IKE Security Association (SA), while IKE\_AUTH, negotiates key material and parameters to establish the VPN tunnel.

One of the main challenges in scaling IPsec VPNs in enterprise networks is their traditional reliance on static point-to-point tunnels. Cisco’s DMVPN addresses this limitation by enabling spoke-to-spoke IPsec tunnels on demand in a full-mesh topology.

Cisco DMVPN is a Cisco IOS Software-based security solution for building scalable enterprise VPNs that support distributed applications such as voice and video.

\begin{figure}[!htb]
	\begin{minipage}{0.48\textwidth}
		\centering
		\includegraphics[width=.98\linewidth]{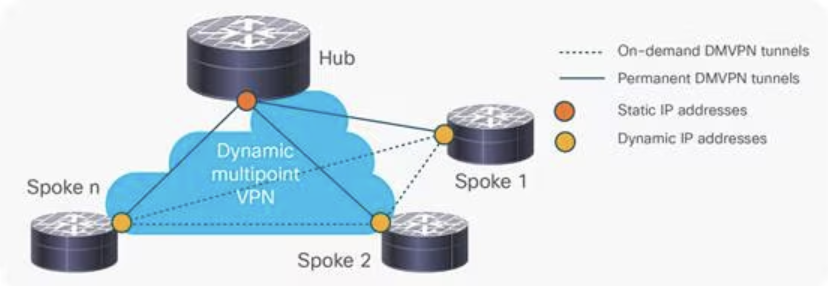}
		\caption{Cisco Dynamic Multipoint VPN (DMVPN) enables spokes to establish permanent IPsec tunnels with a central hub, while also allowing dynamic, on-demand creation of direct IPsec tunnels between spokes without manual configuration. Image from Cisco official documentation.}
		\label{fig:dmvpn-topology}
	\end{minipage}\hfill
	\begin{minipage}{0.48\textwidth}
		\centering
		\includegraphics[width=.98\linewidth]{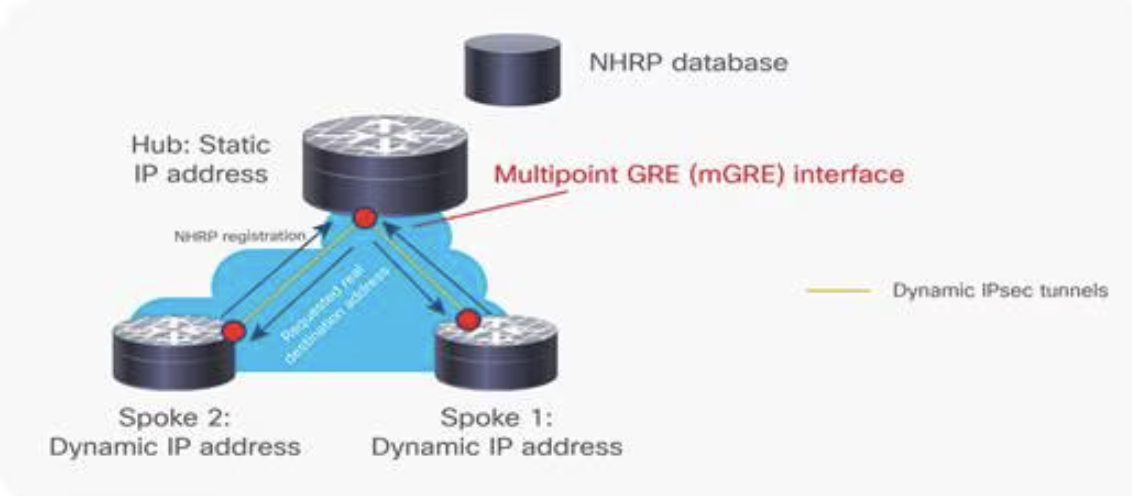}
		\caption{The Cisco DMVPN architecture combines GRE tunneling with IPsec encryption, leveraging the Next Hop Resolution Protocol (NHRP) for dynamic routing and scalable spoke-to-spoke communication. Image from Cisco official documentation.}
		\label{fig:dmvpn-architecture}
	\end{minipage}
\end{figure}

Cisco DMVPN can be deployed in conjunction with Cisco IOS Firewall and Cisco IOS IPS, as well as quality of service (QoS), IP Multicast, split tunneling, and routing-based failover mechanisms. Large-scale, highly available Cisco DMVPN deployments are made possible by load balancing multiple Cisco DMVPN hubs.
Cisco DMVPN can be deployed in two ways:

\begin{itemize}
	\item Hub-and-spoke deployment model: In this traditional topology, remote sites (spokes) are aggregated into a headend VPN device at the corporate headquarters (hub). Traffic from any remote site to other remote sites would need to pass through the headend device. Cisco DMVPN supports dynamic routing, QoS, and IP Multicast while significantly reducing the configuration effort.
	\item Spoke-to-spoke deployment model: Cisco DMVPN allows the creation of a full-mesh VPN, in which traditional hub-and-spoke connectivity is supplemented by dynamically created IPsec tunnels directly between the spokes. With direct spoke-to-spoke tunnels, traffic between remote sites does not need to traverse the hub; this eliminates additional delays and conserves WAN bandwidth. Spoke-to-spoke capability is supported in a single-hub or multihub environment. Multihub deployments provide increased spoke-to-spoke resiliency and redundancy.
\end{itemize}

The 80:20 traffic rule can be used to determine which model to use:

\begin{itemize}
	\item If 80 percent or more of the traffic from the spokes are directed into the hub network itself, deploy the hub-and-spoke model.
	\item If more than 20 percent of the traffic is meant for other spokes, consider the spoke-to-spoke model.
\end{itemize}

Medium-sized and large-scale site-to-site VPN deployments require support for advanced IP network services such as:

\begin{itemize}
	\item IP Multicast: Required for efficient and scalable one-to-many (i.e., Internet broadcast) and many-to-many (i.e., conferencing) communications, and commonly needed by voice, video, and certain data applications
	\item Dynamic routing protocols: Typically required in all but the smallest deployments or wherever static routing is not manageable or optimal
	\item QoS: Mandatory to ensure performance and quality of voice, video, and real-time data applications
\end{itemize}

Traditionally, supporting these services required tunneling IPsec inside protocols such as Generic Route Encapsulation (GRE), which introduced an overlay network, making it complex to set up and manage, and limiting the scalability of the solution. Indeed, traditional IPsec only supports IP Unicast, making it inefficient to deploy applications that involve one-to-many and many-to-many communications.

Cisco DMVPN combines GRE tunneling and IPsec encryption with Next-Hop Resolution Protocol (NHRP) routing in a manner that meets these requirements while reducing the administrative burden.

Key components include:

\begin{itemize}
	\item Multipoint GRE (mGRE) tunnel interface: Allows a single GRE interface to support multiple IPsec tunnels, simplifying the size and complexity of the configuration.
	\item Dynamic discovery of IPsec tunnel endpoints and crypto profiles: Eliminates the need to configure static crypto maps defining every pair of IPsec peers, further simplifying the configuration.
	\item NHRP: Allows spokes to be deployed with dynamically assigned public IP addresses (i.e., behind an ISP’s router). The hub maintains an NHRP database of the public interface addresses of each spoke. Each spoke registers its real address when it boots; when it needs to build direct tunnels with other spokes, it queries the NHRP database for real addresses of the destination spokes. 
\end{itemize}

As topological complexity grows in IPsec infrastructures, especially in dynamic networks, there is an increasing need for solutions that simplify key provisioning without compromising key strength. Cisco's Secure Key Integration Protocol (SKIP) offers a simple solution by allowing the import of Postquantum Preshared Keys (PPKs) from an external key source.  SKIP currently uses HTTPS-based communication over TLS 1.2 with PSK-DHE cipher suites, which can be used to get key material with quantum-safe properties. This interface can be integrated with QKD systems as well as PQC algorithms that act as compliant key sources synchronized between initiator and responder in any node of the network.

\begin{figure*}[h!]
	\centering
	\includegraphics[width=.8\linewidth]{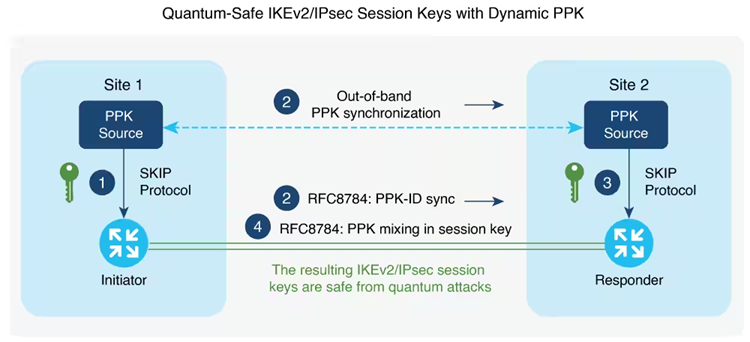}
	\caption{Quantum-Safe IKEv2 and IPsec Session Keys with Dynamic PPK: The IKEv2 initiator and responder are connected to their local key source and configured with the SKIP client that specifies the IP address and port of the key source and the preshared key for the TLS1.2 session. The PPK sources are configured with the SKIP parameters, including the local key source identity and the list of identities of the peer key sources. The following is a high-level operation of the Cisco SKIP protocol: 1) The IKEv2 initiator places a request for a PPK from its key source. The key source replies with a PPK and the corresponding PPK ID. 2) The initiator-side key source synchronizes the PPK to the responder-side key source using an out-of-band mechanism that is specific to the type of key source. The IKEv2 initiator communicates the PPK ID to the IKEv2 responder over IKEv2 using the RFC 8784 extensions. 3) The IKEv2 responder requests from its key source, the PPK corresponding to the PPK ID received from the IKEv2 initiator. The key source replies with the PPK corresponding to the PPK ID. 4) The IKEv2 initiator and responder mix the PPK in the key derivation, as specified in RFC 8784. The resulting IKEv2 and IPsec session keys are quantum-safe. Image from Cisco official documentation.}
	\label{fig:qsikev2}
\end{figure*}

CC protocols such as RSA, DH, and ECDH are the basis of classical secure communication. Secure key exchange and digital signatures have been implemented using algorithms based on RSA, whose security is based on the difficulty of factoring large integers, or on DH/ECDH, whose computational complexity derives from the discrete logarithm problem in finite groups or over elliptic curves. In spite of their historical effectiveness, these mechanisms are vulnerable to quantum algorithms like Shor's~\cite{shor}, which can solve the underlying mathematical problems in polynomial time, compromising the long-term security of encrypted data. 

In this context, QKD emerges as a promising technology achieving information-theoretic security. With prepare-and-measure QKD protocols, keys are modulated in light degrees of freedom by a transmitter and sent through a quantum channel to a receiver, where eavesdropping results in increased excess noise for CV-QKD or in Quantum Bit Error Rates (QBER) for DV-QKD. Those noises can be detected and quantified and provide an upper limit on the information which might have leaked to the eavesdropper. The precise relationship between excess noise or QBER and the amount of leaked information is provided by the relevant security proofs in the finite size regime~\cite{leverrier_finite-size_2010, Lim_Curty_2014}. In order to distill a symmetric secure key, the parties perform error correction and privacy amplification, which lowers the key rate and reduces the information leaked to a chosen low value.

PQC offers algorithmic defense against quantum attacks, especially when combined with QKD. The NIST ML-KEM algorithm is, currently, the only standardized PQC option for key agreement. The most widely used version in this experiment, ML-KEM-1024, is the most secure of the three standardized ML-KEM flavors with Level 5~\cite{FIPS203}.

The combination of ML-KEM and QKD in a unified hybrid framework results in a new security paradigm. Since both methods are quantum-safe, this model combines the hardware based security model of QKD alongside the freedom of software integration offered by ML-KEM. Within this multi-location VPN topology, QKD ensures key distribution with detection of a possible eavesdropper, while ML-KEM fortifies the key encapsulation process using lattice-based security.

This layered approach ensures cryptoagility and defense in depth. It extends the lifecycle of classical cryptography by embedding PQC safeguards and introduces QKD for critical links where absolute confidentiality is required and the fiber reach and loss allow for it. Such hybrid schemes are particularly valuable in banking environments, where the compromise of cryptographic keys could lead to severe financial and reputational losses. The seamless integration of these technologies, enabled through SDN-based management and following standards protocols and interfaces, provides a robust blueprint for secure, quantum-resilient financial communication networks.

\section{State of the Art}

The financial sector is increasingly recognizing the imminent threat posed by quantum computing to classical cryptographic systems. In response, institutions are actively exploring and implementing quantum-safe solutions, QKD as well as PQC, to safeguard sensitive data and maintain trust in digital transactions.

QKD has transitioned from theoretical research to practical applications within the banking industry. Notably, in 2004, the world's first bank transfer secured by QKD was executed in Vienna, Austria~\cite{peev2009secoqc}, marking a significant milestone in quantum-secure financial transactions. More recently, JPMorgan Chase demonstrated a 100 Gbps quantum-safe IPsec VPN tunnel over a 46 km deployed fiber, showcasing the feasibility of integrating QKD into high-speed banking networks~\cite{alia2024100}. Also JPMorgan Chase in \cite{pistoia2023paving} demonstrates a 800 Gbps quantum-secured channel, carrying data encrypted with QKD at distances up to 100Km. Both JPMorgan PoCs were executed using a single QKD link, relying on equipment from a single vendor throughout the entire setup.

HSBC pioneered the use of QKD to secure AI-powered foreign exchange trading, successfully protecting a simulated €30 million EUR/USD transaction~\cite{HSBC_1}. This trial was conducted over the UK's Quantum-Secured Metro Network, linking HSBC's Canary Wharf headquarters with a data center in Berkshire. In addition, HSBC has explored the application of PQC to safeguard digital assets. The bank tested the use of PQC algorithms to protect tokenized physical gold transactions on its Orion platform, ensuring resilience against future quantum attacks~\cite{HSBC_2}. HSBC has collaborated with Quantinuum to investigate quantum-enhanced cybersecurity measures, including the integration of quantum-generated cryptographic keys to reinforce defenses against advanced cyber threats. Similarly, Toshiba and SoftBank successfully completed a field experiment of an IPsec QKD-VPN, demonstrating a practical implementation in inter-site communications~\cite{SoftBank}. 

Recognizing the complementary strengths of QKD and PQC, some organizations advocate for a hybrid approach~\cite{ETSI-QSC-Report-2025}. For instance, ID Quantique~\cite{IDQ_defenceindepth} promotes a dual strategy for quantum-safety that leverages the physical security of QKD and the algorithmic robustness of PQC, aiming for a layered defense-in-depth model suitable for critical financial operations. In such approach, the risk of Harvest Now, Decrypt Later (HNDL) attack is mitigated early, usually between data centers and clusters where the surface of the attack is the greatest. It also allows for greater flexibility for the migration of the network infrastructure to PQC, as well as a higher level of security assurance should there be a need to replace PQC algorithms down the road due to newly discovered vulnerabilities.

Integrating quantum-safe technologies with existing protocols such as IPsec is crucial to seamless adoption. In recent years, various approaches have been proposed to integrate Quantum Key Distribution (QKD) with IPsec. A common line of work focuses on embedding QKD directly into the IPsec core architecture~\cite{marksteiner2018resilience}, where the design considers the characteristics and key generation rates of QKD devices to optimize their utilization within the IPsec framework. Other efforts have explored the integration of IPsec with modern network paradigms such as SDN. For instance, Heydari et al.~\cite{heydari2014integrating} proposed the embedding of IPsec functionality into SDN following OpenFlow models, while Yunchun and Jutao~\cite{li2015sdn} developed a method to manage IPsec-specific constraints using OpenFlow switches in conjunction with SDN controllers. More recently, the Internet Engineering Task Force (IETF) approved an Request For Comments (RFC)~\cite{marin2021yang} defining a YANG data model~\cite{bjorklund2016yang} for full IPsec management within SDN architectures. Building on this model, the work presented by Ruben et al.~\cite{b.mendezQuantumResistantSoftware2026} demonstrates how it can be leveraged to facilitate the integration of QKD into IPsec through SDN orchestration. Several vendors like ADVA, Rohde \& Schwarz and Fortinet have developed solutions enabling the retrieval of IPsec keys using QKD systems through the standardized API of ETSI, ETSI GS QKD 004~\cite{ETSI_004} and ETSI GS QKD 014~\cite{ETSI_014} facilitating the incorporation of quantum security into established VPN infrastructures. Standardization bodies are actively working to ensure interoperability and security in quantum-safe implementations. IETF is currently standardizing the support for ML-KEM in IKEv2~\cite{ietf-ipsecme-ikev2-mlkem-03}. In the QKDN domain, ITU-T Y.3800~\cite{ITU-T_Y.3800} offers an overview of networks supporting QKD, Y.3801~\cite{ITU-T_Y.3801}  defines functional requirements across the quantum, key management, control and management layers, Y.3802~\cite{ITU-T_Y.3802}  specifies the functional architecture of QKDNs, and Y.3803~\cite{ITU-T_Y.3803}  focuses on key management procedures. ITU-T also has proposed frameworks for integrating QKD with IPsec~\cite{ITU-T_Y.QKD-IPSec-fr}, aiming to combine quantum key distribution's secure key exchange with IPsec's robust encryption mechanisms. In parallel, ETSI GS QKD 004 and ETSI GS QKD 014 define interoperable key-delivery interfaces toward applications, while ETSI GS QKD 015 specifies the control interface for SDN-enabled QKD nodes. From this perspective, our proposal should be understood as standards-aligned rather than fully standards-native: it reuses ETSI-compatible interfaces and SDN abstractions where possible, but it also incorporates operational elements such as Cisco SKIP and DMVPN in order to interoperate with currently deployed enterprise IPsec infrastructures.



International collaborations are accelerating the deployment of quantum-safe networks. Singapore's National Quantum-Safe Network Plus (NQSN+) initiative supports nationwide deployment of quantum-safe solutions, including QKD and PQC, providing businesses with access to advanced security infrastructures~\cite{singapour}. Similarly, a Luxembourg-led partnership is implementing cross-border QKD networks, highlighting the global momentum in securing financial communications~\cite{luxemb}. A relevant initiative in this context is the Madrid Quantum Communication Infrastructure (MadQCI)~\cite{MadQCI}, developed as part of Spain’s contribution to the EuroQCI program. MadQCI integrates QKD systems with classical network infrastructure to build a quantum-secure network in the Madrid region. It serves as a reference for interoperability and hybrid quantum-classical deployments, reinforcing the use of QKD in real-world network communication.

Despite these advancements, challenges persist in integrating quantum-safe technologies into existing banking infrastructures. The complexity of current systems, the need for interoperability among diverse vendors, and the requirement for minimal disruption to services necessitate careful planning and phased implementation strategies. Authorities like the UK's National Cyber Security Centre (NCSC) have outlined timelines for transitioning to post-quantum cryptography, recommending organizations to identify critical services for upgrade by 2028 and complete the transition by 2035~\cite{ncsc}. The European Commission has published a coordinated roadmap for the transition to PQC, aimed at public administration entities and other critical
infrastructures, recommending the PQC transition planning by the end of 2026, completing the transition of high-risk use cases by the end of 2030, and the transition of medium-risk use cases by 2035~\cite{Timeline_Post_Quantum}. Such roadmaps are essential to guide financial institutions through the complex migration process, as indicated  by FS-ISAC, the Quantum Safe Financial Forum and members of the Canadian Forum for Digital Infrastructure Resilience~\cite{Implementation_Roadmap}.

\section{Methods}
\subsection{Architecture}
The architecture is built on the key distribution framework, which has a system of distributed KMSs, regulated by a central controller. This system acts as a central point in coordinating key management, identifying the type of key that can be established between two nodes. It uses the maximum security guarantees between any pair of nodes in the network. The controller is capable of managing path calculation, ensuring the correct distribution of keys between non-neighbouring nodes and the allocation of the key delivery to each service that requires it.

The key generation side consists of PQC modules that are part of the key management system to make up for the lack of PQC support at the endpoints; and by pairs of QKD nodes wherever possible, depending on the distance and quality of the fibre.

The consumer side consists of IPSEC endpoints capable of requesting an external symmetric key and negotiating protected tunnels with it, for instance, based on RFC8784. Key delivery is performed using ETSI GS QKD 004 and ETSI GS QKD 014 interfaces from the QKD layer, and Cisco SKIP from KMS layer. In this system, QKD is used to generate and distribute high-entropy symmetric keys combining with classical and PQC key encapsulation mechanisms. As a result, the overall system follows a hybrid security model, with access to a continuous stream of fresh, independent quantum-safe keys, that increase the security using frequent rekeying, while shortens each key’s exposure window. Our method substantially strengthens the overall posture relative to purely classical approaches by combining quantum-safe key material (QKD+PQC) with continuously refreshed symmetric encryption aligned to operational requirements.

\subsection{Network Architecture}
\begin{figure*}[ht!]
	\centering
	\includegraphics[width=.95\linewidth]{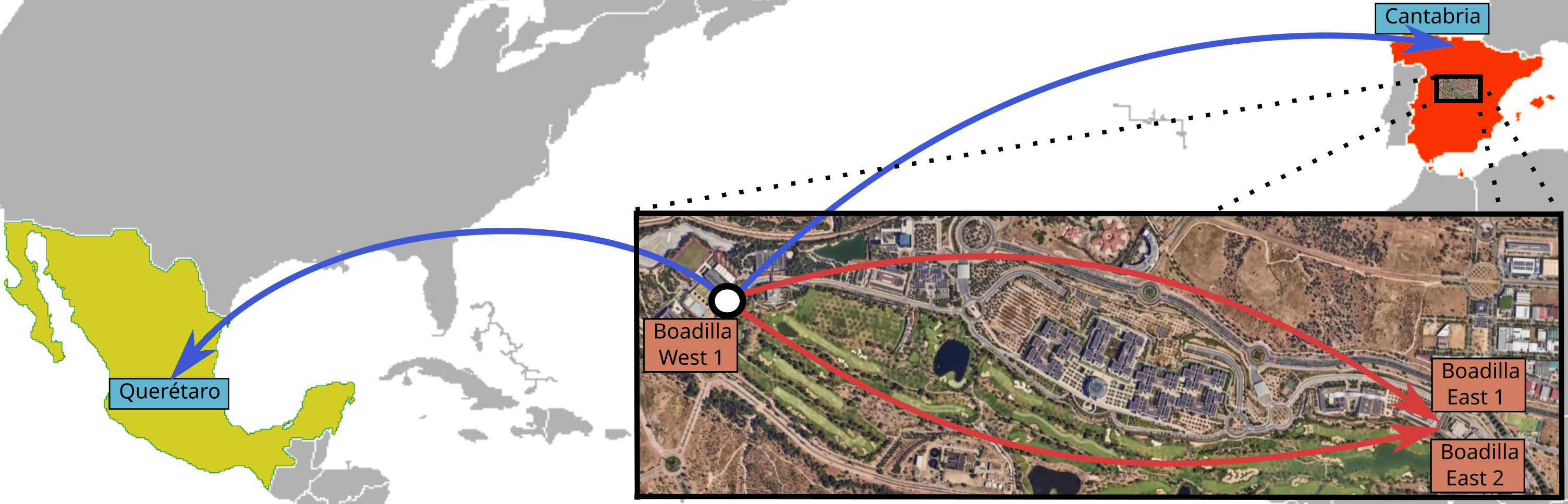}
	\caption{This figure illustrates the physical layout of the experimental quantum-safe network deployed for this study. At the center of the topology is the hub node located in Madrid (Boadilla West 1), which hosts the SDN controller, as well as direct access to two QKD links, one provided by LuxQuanta and the other by ID Quantique. These QKD links interconnect the three nodes in the Madrid metropolitan area, two of them collocated (Boadilla East 1 and Boadilla East 2) forming the Quantum Trust Domain. The figure also shows the PQC-based connections that extend from Madrid to two cloud nodes: one located in Cantabria (northern Spain) and the other in Querétaro (México). These remote links are secured through PQC tunnels using ML-KEM. The hybrid architecture, combining quantum and post-quantum links, is visually represented through the coexistence of fiber-based QKD channels and IP based PQC links.}
	\label{fig:FullScheme}
\end{figure*}

The experimental quantum-safe communication infrastructure presented in this work was designed to mirror a realistic enterprise-grade financial network, integrating advanced cryptographic protocols into legacy infrastructure without compromising operational integrity. The network consists of five interconnected nodes organized logically into a star topology, with the central hub located at Banco Santander's facilities in Ciudad Financiera in Madrid region, the physical deployment can be seen on Figure \ref{fig:FullScheme}. This central hub is responsible for orchestrating cryptographic material provisioning to all the nodes arranged around and connected to the central node.

The QKD setup comprises the three physical nodes located in Madrid, each integrated with two independent vendors (IdQuantique and Luxquanta), this deployment enables resilience by mitigating risks associated with potential vulnerabilities in a single-vendor technology stack and by hybridizing DV-QKD and CV-QKD keys. Moreover, it validates the vendor agnostic nature of the architecture, made possible by the SDN controller adhering to ETSI GS QKD interface standards. Despite differences in QKD protocol implementations, all nodes operate correctly, underscoring the effectiveness of using standardized APIs and protocols within a centrally managed SDN framework. Nodes within QKD reach will benefit from QKD-derived key material, increasing the baseline protections derived from the use of PQC-only of the complete system.

The PQC setup is enabled on all five nodes, with the two nodes deployed in the Santander's private cloud environment in Cantabria (Spain) and Querétaro (México) limited to PQC-only communications. The Key Managememnt Service (KMS) instances negotiate keys among them using ML-KEM1024 with any other node in the network. This allows for any-to-any communication independent from the distance and direct fiber availability.

Each node in the deployment comprises a set of coordinated subsystems: a set of Cisco routers configured for dynamic VPN negotiation and SKIP-based key ingestion, more specifically 3 Cisco Catalyst 8300 have been used in the Madrid locations, and 2 Cisco Catalyst 8300V for the nodes in the Cloud. 

The following excerpt shown on Figure \ref{fig:IPseconf} illustrates a representative configuration snippet from a Cisco router enabling dynamic PPK within an IPsec tunnel establishment process. In this setup, the router is configured to retrieve symmetric key material from an external Key Management System module, potentially integrated with a QKD device. The PPK is dynamically injected as part of the Security Association (SA) establishment, as per RFC 8784~\cite{rfc8784}, providing quantum safety to the negotiated keys~\cite{ciscoppk}. When the endpoints are Cisco IOS XE devices, the PPKs are also dynamically injected as part of the SA rekeying, against what RFC 8784~\cite{rfc8784} mandates, to provide quantum-safe forward secrecy. 

\begin{figure}[h!]
	\centering
	\begin{lstlisting}[language=bash,basicstyle=\ttfamily\footnotesize,frame=single]
		conf t
		hostname router-be1
		!
		no crypto ikev2 proposal default
		crypto ikev2 proposal ikev2-prop
		encryption aes-gcm-256
		prf sha512
		group 20
		!
		no crypto ikev2 policy default
		crypto ikev2 policy ikev2-pol
		match fvrf any
		proposal ikev2-prop
		!
		crypto ikev2 keyring psk-keyring
		peer all
		address 0.0.0.0 0.0.0.0
		pre-shared-key xxxxxxxxxxxxxxxx
		!
		!
		crypto ikev2 keyring ppk-keyring
		peer all
		address 0.0.0.0 0.0.0.0
		ppk dynamic skip-client Required
		!
		!
		crypto ikev2 profile ikev2-quantum
		match identity remote any
		authentication remote pre-share
		authentication local pre-share
		keyring ppk ppk-keyring
		keyring local psk-keyring
		lifetime 600
		dpd 10 2 periodic
		!
		crypto sks-client skip-client
		server ipv4 x.x.x.x port xxxxxxx
		psk id xxxxxx key xxxxxxxxxxxxxxxx
		!
		crypto ipsec transform-set esp-gcm256 esp-gcm 256
		mode transport
		!
		crypto ipsec profile ipsec-profile-quantum
		set transform-set esp-gcm256
		set ikev2-profile ikev2-quantum
		!
		interface Tunnelx
		vrf forwarding quantum
		ip address x.x.x.x 255.255.255.0
		no ip redirects
		ip nhrp map multicast x.x.x.x
		ip nhrp network-id xxx
		ip nhrp nhs x.x.x.x nbma x.x.x.x
		tunnel source GigabitEthernet0/0/4.x
		tunnel mode gre multipoint
		tunnel protection ipsec profile ipsec-profile-quantum
		!
	\end{lstlisting}
	\caption{The specific IP addresses, ports and other specific details in the code above are provided as an example only, and not reflecting actual configuration of the devices used in this project.}
	\label{fig:IPseconf}
\end{figure}

A local telemetry system based on syslogs is used for registering all the events. The three Madrid-based nodes are positioned in different Santander facilities and interconnected via metropolitan fiber capable of supporting quantum channels. These nodes were connected via point-to-point QKD links. It is important to say that no new fibers were deployed: we used pre-existing generic production fibers between Santander Bank datacenters.

Each Trusted Node in the network is equipped with a localized SDN stack, which includes an SDN agent functioning as the node's local control element and a dedicated Local Key Management System (LKMS) responsible for the acquisition, storage, and delivery of cryptographic key material. This internal architecture adheres to the ETSI GS QKD 015~\cite{ETSI_015} standard, which defines the functional interface between SDN-QKD nodes and the logically centralized SDN controller. The SDN agent ensures that each node can interact autonomously with the central controller, and the KMS guarantees that quantum and post-quantum keying material is securely provisioned to connected encryption devices.

\subsection{QKD Infrastructure and Integration}
The Quantum Key Distribution infrastructure is deployed within the Madrid region segment of the network, forming what we refer to as the \textit{Quantum Trust Domain}. This domain includes three sites, in two geographically distinct locations, each interconnected via dedicated optical fiber links capable of supporting both quantum and classical channels.

Each QKD system generates a continuous pool of key material, which is buffered in the LKMS of each Trusted Node. Keys are tagged with metadata such as key ID, originating QKD link, and expiration time. This allows the SDN controller to assign keys to specific sessions, IPsec tunnels, or applications according to predefined security policies.

The chosen QKD technology combines both CV-QKD and DV-QKD devices. Two distinct QKD vendors are integrated into this architecture: LuxQuanta and ID Quantique. Their simultaneous operation within the same QKD domain allows for the evaluation of multivendor, multi-technology integration in a single network.

The quantum links are laid over standard SMF-28 G.652.D single-mode fiber and support both quantum and classical channels using either time-division or wavelength-division multiplexing. CV-QKD technology is used in this deployment due to its compatibility with conventional optical hardware and operational tolerance in metro scale networks. Both QKD systems are configured in a prepare-and-measure setup where the transmitter encodes the data into the states of light, while the receiver measures the states. The details of the modulation and measurement processes are the main aspects that distinguish CV-QKD  from DV-QKD. In LuxQuanta's CV-QKD system, the transmitted data are modulated into the quadratures of light using Gaussian modulated coherent states and measured at the receiver using homodyne coherent detection. These links are designed to provide high key generation rates (in the order of kbps) suitable for symmetric key exchange in high-throughput IPsec tunnels. ID Quantique’s DV-QKD systems operate using weak coherent pulses transmitted over standard single-mode optical fiber. The sender unit employs phase and time-bin modulation to encode bits onto single-photon level light pulses. On the receiver side, time-bin detection is used to obtain the data bits, while interferometric detection tests for the presence of an eavesdropper. Both rely on single-photon detectors, often based on avalanche photodiodes or on superconducting nanowire single-photon detectors (SNSPD’s) - to measure the quantum states.

Each QKD link is paired with an authenticated classical channel that carries reconciliation, privacy amplification, and metadata coordination messages between emitter and receiver. These channels also support SDN control messages for tunnel instantiation and key allocation. Importantly, the architecture adheres to the ETSI GS QKD 004 and 014 standards for secure QKD key delivery to KMS modules, to ensure modular integration and vendor interoperability. Both interfaces are used in a complete transparent manner thanks to the SDN layer governing the experiment.

\begin{figure*}[h!]
	\centering
	\includegraphics[width=1\linewidth]{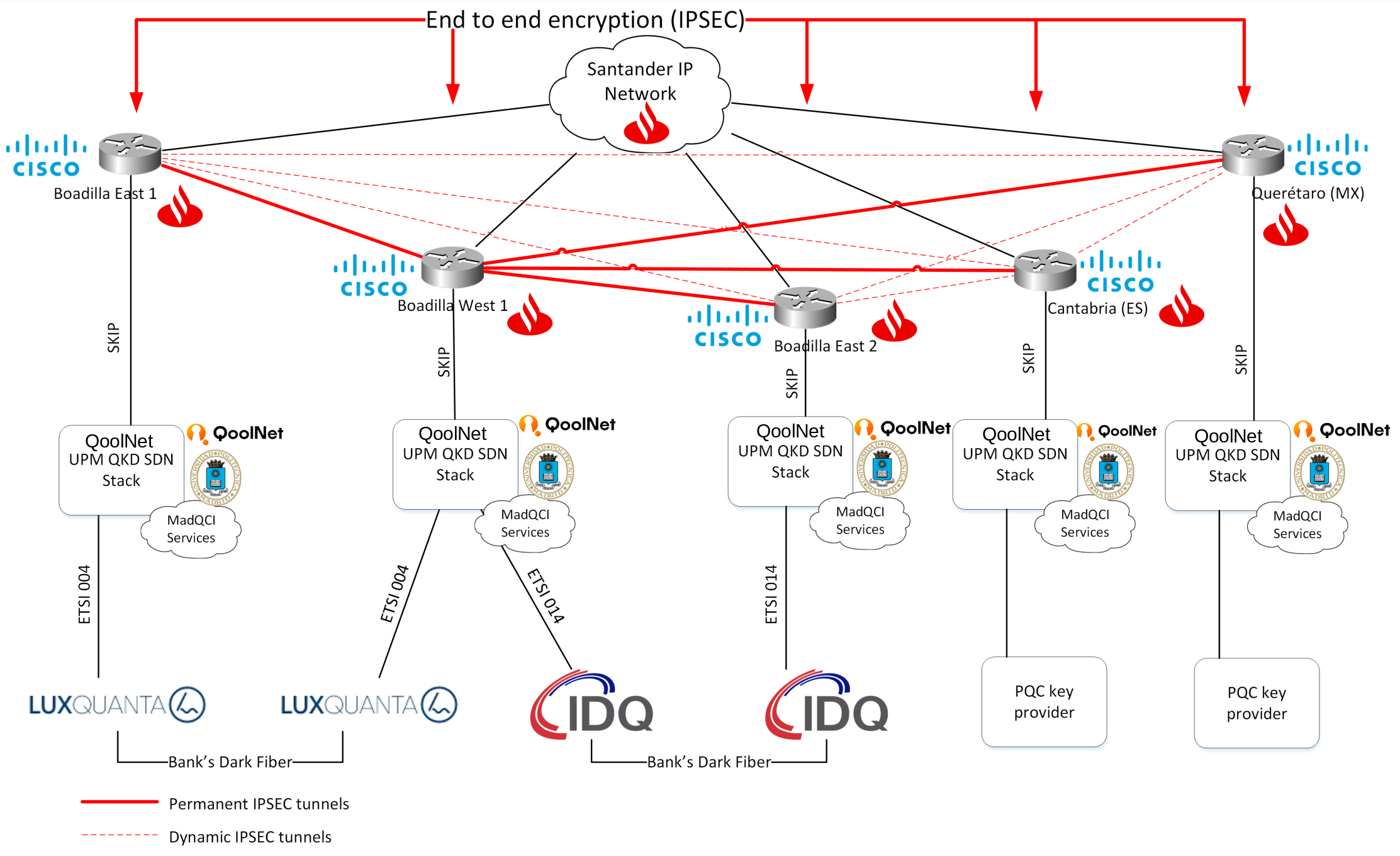}
	\caption{High-level schematic overview of the quantum-safe network architecture deployed in the experiment. The diagram captures all five participating Trusted Nodes: three physically connected with QKD links within the Madrid region and two remote endpoints located in Cantabria and Querétaro. In the lower section of the figure, the underlying communication infrastructure is depicted, distinguishing between the quantum trust domain—secured by QKD links provided by LuxQuanta and ID Quantique and the PQC links that connect to Cantabria and México. These PQC links, secured using ML-KEM, illustrate the network’s capacity to maintain quantum-safe links across geographic and trust boundaries. Above this infrastructure layer, the SDN plane is shown, it manages all key lifecycle operations, routing and secure distribution of key material between end nodes. At the top most layer, the Cisco routers are illustrated, each equipped with SKIP interfaces to receive keys and configured to support both permanent and on-demand IPsec tunnels. The figure highlights the DMVPN configuration that enables dynamic full-mesh connectivity across the network, allowing to establish secure tunnels as required by the application layer, triggered by the Cisco routers.}
	\label{fig:SDNIPsecQKD}
\end{figure*}

The remote nodes, located in Cantabria and Quer\'etaro, reside outside the Quantum Trust Domain and are instead connected using PQC techniques. These nodes simulate real-world financial branch offices or cloud-edge environments that may not be capable of supporting QKD due to infrastructure constraints like losses. This bifurcation of the network into quantum and post-quantum zones allows for evaluation of heterogeneous cryptographic policy enforcement and dynamic interoperation between trust domains. The deployment scheme can be seen on Figure \ref{fig:SDNIPsecQKD}

\subsection{PQC Configuration and Algorithms}
PQC forms the foundational security layer for all communication links outside the Quantum Trust Domain. The selected algorithm ML-KEM for key encapsulation is a NIST standardization winner and provides quantum-resilient security grounded in lattice-based hardness assumptions.

ML-KEM is implemented to offer Quantum Safe communications where QKD is not possible due to distances and losses. The algorithm operates efficiently between the Trusted Nodes and integrates into the key delivery pipeline via SKIP, maintaining modularity between the key generation and transport layers. PQC operations are implemented using the Open Quantum Safe (liboqs) library~\cite{oqs2025liboqs}, compiled and statically linked into the SDN framework.

The PQC key lifecycle is managed locally by each LKMS couple, which monitors key usage rates, expiration thresholds of each session, and the success rates of key exchange operations. Keys are pre-provisioned for rapid delivery to the application layer (Cisco routers in this case), and keys are internally tagged with metadata including purpose and expiration time.

\subsection{Hybrid Key Exchange Method}
Hybrid cryptographic mechanisms aim to combine multiple key exchange techniques, typically one classical (such as ECDH) and one quantum-resistant, QKD and/or post-quantum KEM, in such a way that the security of the resulting session key depends on the security of both methods. This approach ensures that even if one mechanism is broken (e.g., ECDH broken by a quantum computer), the session remains secure due to the strength of the second algorithm. The encryption layer used in this field trial natively uses the ECDH Group 20 in the context of IKEv2. It refers to ECDH over the P-384 curve, as specified in RFC 5903. This group provides fast key agreement and strong classical security, and is widely adopted in secure communication protocols.

In the hybrid method described in RFC 8784 the resulting session key is derived from both the DH/ECDH exchange and a PPK injected into the IKEv2 key derivation process to provide \textbf{post-quantum resistance}. In this field trial, the PPK is provided through both QKD and PQC. The steps involved are as follows:

\begin{enumerate}
	\item During the \texttt{IKE\_SA\_INIT} exchange, the two peers perform a standard ECDH exchange (using Group 20 in our setup), resulting in a shared secret.
	
	\item According to the RFC 8784~\cite{rfc8784}, Section 3, the value of the key SKd is derived among other key values but this key is not used right away and referenced in the next steps as \text{SK}'d)
	
	\item As an additional step, another key, the PPK is delivered out-of-band through Quantum Safe methods, PQC (using ML-KEM1024 in our setup) and/or QKD
	
	\item The final session key SKd is derived by mixing both values using a pseudorandom function (PRF): 
	\[
	\text{SK}_d = \text{PRF}(\text{PPK}, \text{SK}'d)
	\]
\end{enumerate}

This combination ensures that even if ECDH is broken by a quantum computer, the attacker would still need to know the PPK to derive the session key, thus achieving resistance against ”store/harvest now, decrypt later” (SNDL/HNDL) attacks. This hybrid approach is particularly suitable for transitional post-quantum security in IPsec deployments while waiting for full standardization and adoption of NIST post-quantum KEM algorithms. The solution maintains interoperability and high performance by using ECDH Group 20 while incorporating future-proof security enhancements via PPK or quantum-safe keys.

The routers can be configured to require the use of PPKs or to support them as an optional feature, facilitating deployment in existing networks by enabling a transitional phase with mixed RFC 8784 support. In our setup, all sessions required the use of a PPK; without it, the IKEv2 SA establishment would not complete.

\subsection{SKIP Protocol}
The Simple Key Management for Internet Protocol (SKIP) provides an interface between the Trusted Nodes and the distributed router endpoints. The SKIP interface allows routers to request key material without needing to understand the source or algorithm used to generate QKD-derived symmetric keys, PQC encapsulated keys, or traditional pre-shared secrets.

SKIP operates over HTTPS and each transaction includes a key request from a router, SDN controller side evaluation, and secure delivery of key material. The delivered keys include metadata indicating usage constraints, expiration time, and link-specific security attributes. As of the time in which the experiment was executed, the Cisco devices did not support ML-KEM in the SKIP interface. Instead, to enable quantum-safety in this interface, these TLS sessions used a preshared key. It is important to remind, though, that the device to SKIP endpoint communication is strictly local within the trusted perimeter, hence limiting any exposure to third parties. 

The architecture supports on-demand rekeying and cryptographic failover. If a QKD channel degrades or key buffer levels fall below policy thresholds, the SDN controller switches to PQC-generated keys.

SKIP also supports key lifecycle logging and auditing. The LKMS enforces cryptographic isolation, ensuring that key material is never reused outside its defined scope.

\subsection{IPsec Tunnel Establishment and DMVPN Behavior}
IPsec is used as the underlying protocol suite for providing confidentiality, integrity, and authentication between nodes. Each router participates in a Cisco DMVPN configuration enabling dynamic formation of IPsec tunnels. The DMVPN model is distributed between the Madrid Quantum Trust Domain and the cloud virtual routers at Cantabria and Querétaro.

When secure communication is triggered, the initiating router sends a key request to their local Trusted Node (throught the LKMS), the destination router makes the same operation on the other side of the secure communication, when both requests are received by the SDN controller, it selects the appropriate paths to deliver symmetric keys between the endpoints.

Tunnels are monitored for data rate, error rate, and expiration, locally on each router. Rekeying is triggered based on expiration policies and performed without disrupting sessions. DMVPN configuration also supports application-specific routing and tunnel hardening.

The architecture supports flexible tunnel management, allowing for application-driven security policies. With this approach, the key material is delivered to the routers independently of the quantum safe source, that being QKD or PQC.

\subsection{SDN Architecture}

\begin{figure}[h!]
	\centering
	\includegraphics[width=.78\linewidth]{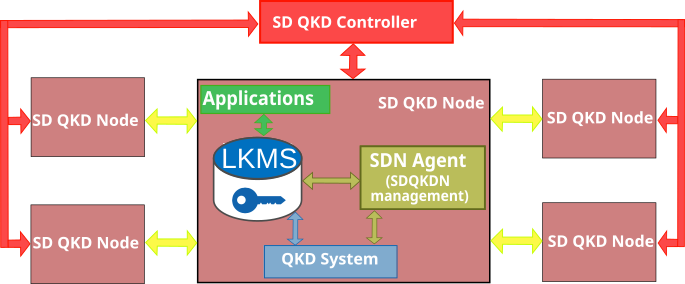}
	\caption{The figure shows the SDN architeture used in the experiment, it shows the representative nodes in the network, each interconnected through a combination of quantum and classical channels. These include the PQC links, the QKD links (that includes quantum channels for transmitting quantum signals, service channels required for maintaining quantum link stability and classical channels for key distillation processes), and data channels for user traffic. Additional classical communication paths may exist between encryption devices or software-based applications operating within a node. It is important to note that the hub node (Boadilla West 1) serves as releay node when the other nodes, physical or cloud, requires an E2E key.}
	\label{fig:arquitect}
\end{figure}

The proposed architecture is primarily grounded in ETSI GS QKD 015 ~\cite{ETSI_015}, which defines the abstraction and control interface between SD-QKD nodes and higher-level orchestration entities such as SDN controllers. This standard provides the necessary models for resource discovery, capability exposure, application registrations, and key management interaction. It enables a vendor-agnostic and programmable control plane. The use of ETSI GS QKD 015 as the core integration layer allows the SDN controller to manage cryptographic operations, key delivery request and tunnel creation. It is implemented as a modular service, communicating with Trusted Nodes through a local control element present on each node and named as Agent. The SDN controller managed key delivery routers via standardized APIs, primarily based on ETSI GS QKD 004 and ETSI GS QKD 014 for key delivery and ETSI GS QKD 015 for the communication between each node with the SDN controller. The SDN arquitecture proposed on this work can be seen on Figure \ref{fig:arquitect}. The controller maintains a global view of the network, including QKD link health, PQC availability, and tunnel status. The controller enables cryptoagility by adapting operations in response to telemetry. For example, it allows to detect QKD channel degradation and triggers PQC alternative links between the nodes. The controller also provides analytics and auditing, all cryptographic events are logged, not only in the SDN controller but also on any component. The SDN controller thus serves as both the operational and governance element of the quantum-safe infrastructure. Our proposal is conceptually aligned with these ITU-T Y.3803~\cite{ITU-T_Y.3803}, Y.3804~\cite{ITU-T_Y.3804}, Y.3805~\cite{ITU-T_Y.3805} and X.1712~\cite{ITU-T_X.1712} recommendations, particularly in terms of layered architecture, trusted-node operation, and separation between key management and application planes. Finally, ETSI GS QKD 020, which addresses specific aspects of QKD network integration and key management interactions, is not directly applied in this architecture. This is because the interconnection between QKD domains and services is handled transparently through the SDN managed KMS layer, which abstracts the underlying complexity of key transport and relay.

An essential component of the SDN-based architecture is the LKMS, which acts as the central cryptographic manager within each Trusted Node. The LKMS is responsible not only for acquiring key material from the underlying quantum and post-quantum sources (QKD devices and PQC key generation) but also for securely managing the full lifecycle of these keys. Its responsibilities include E2E key transport, ensuring that cryptographic material is securely and reliably delivered to the correct pair of routers participating in an IPsec tunnel. Through its integration with the SDN control plane and the SKIP protocol, the LKMS facilitates dynamic key distribution based on application level demands.

On top of the LKMS is running the SKIP interfaces to connect with Cisco routers as clients of Quantum Safe key material and E2E IPsec tunnel generators. A precise understanding of IPsec tunnel establishment via the SKIP interface is critical. The process starts when a router receives an event requiring an IPSEC tunnel to be established with another node. That router—designated as the Initiator—issues a get\_capabilities request to the local SKIP interface embedded within its corresponding Trusted Node, that in the context of SDN networks it is usually named as SD-QKD node. Upon receiving the capabilities response from this node, the Initiator commences a cryptographic handshake with a peer Cisco router, designated as the Responder. Concurrently, the Responder queries its own local SKIP interface for capability details and, once confirmed, replies to the Initiator to indicate readiness for key exchange.

\begin{figure*}[htp!]
	\centering
	\includegraphics[width=.9\linewidth]{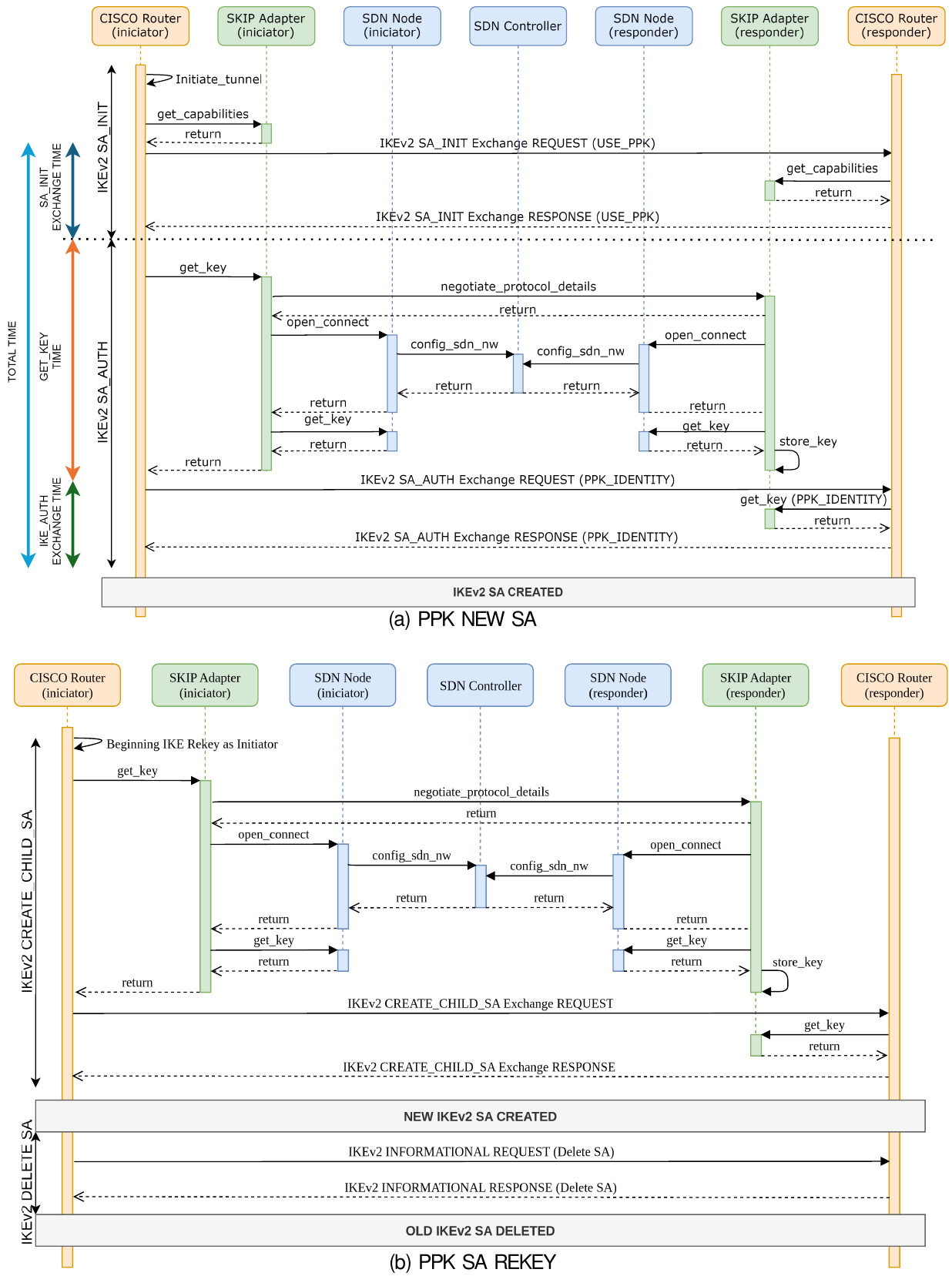}
	\label{fig:figsksperfA}
	
	\caption{The IKEv2 SA setup via the SKIP interface begins when the Initiator router requests cryptographic capabilities from its local SD-QKD Node and initiates a handshake with the Responder router. The Responder also retrieves its capabilities, confirming readiness to proceed. The Initiator then requests specific key material via SKIP, compliant with ETSI GS QKD 004 interface, and shares the key identifier (key\_id) with the Responder, which then independently retrieves the same key from its local node. Ultimately, identical symmetric keys are securely delivered to both routers, enabling the IKEv2 SA creation.}
	\label{fig:skipsteps}
\end{figure*}

Subsequently, the Initiator router explicitly requests key material through its SKIP interface. This request is executed following the specifications defined in ETSI GS QKD 004, which standardizes the interactions between QKD devices and the key delivery infrastructure. The SKIP interface at the Initiator’s SD-QKD node then prepares the appropriate cryptographic key material for delivery. To comply with the SKIP protocol, the Initiator transmits a specific key\_id identifying the required symmetric key to the Responder router. Upon receiving this identifier, the Responder router similarly requests the corresponding key material from its local SD-QKD node. As a result of this coordinated process, the same symmetric cryptographic key is securely provisioned and synchronized on both Initiator and Responder routers, thus enabling the IPsec tunnel to be established with cryptographic assurance and consistency. All this logic is summarized on Figure \ref{fig:skipsteps} (a).

Before the established SA expires, the initiator router initiates a SA rekey repeating the same requests to the SKIP provider. This behavior is against RFC 8784, which mandates that PPKs are not requested on rekey phases. This out-of-standard request will only happen when the session has been established with another Cisco IOS XE endpoint to ensure compatibility, and improves RFC 8784 by providing quantum-safe perfect forward secrecy. This logic is summarized on Figure \ref{fig:skipsteps} (b).

\section{Results}

The implementation of a hybrid QKD and PQC-based network architecture demonstrated both high performance and enhanced security. By leveraging Cisco's DMVPN technology in combination with SKIP interfaces for dynamic key exchange, we successfully established E2E, on-demand encrypted tunnels between nodes in Madrid, Cantabria, and Querétaro. The network's full mesh topology, physically supported by QKD and PQC connections, enabled efficient communication flows and minimized latency, offering flexibility for future expansion. In this section, we analyze the results obtained.

\begin{figure*}[h!]
	\centering
	\begin{lstlisting}[language=bash,basicstyle=\ttfamily\footnotesize,frame=single]
		qkd@node-be1:~$ traceroute NODE-BE2
		traceroute to NODE-BE2(z.z.z.z), 30 hops max, 60 byte packets
		1  ROUTER-BE1 (w.w.w.w)  0.224 ms  0.167 ms  0.135 ms
		2  ROUTER-BW1 (w.w.w.w)  0.741 ms  0.711 ms  0.682 ms
		3  ROUTER-BE2 (w.w.w.w)  1.271 ms  1.241 ms  1.211 ms
		4  NODE-BE2 (z.z.z.z)  1.307 ms  1.278 ms  1.247 ms
		
		qkd@node-be1:~$ traceroute NODE-BE2
		traceroute to NODE-BE2 (z.z.z.z), 30 hops max, 60 byte packets
		1  ROUTER-BE1 (w.w.w.w)  0.224 ms  0.170 ms  0.138 ms
		2  ROUTER-BE2 (w.w.w.w)  0.697 ms  0.637 ms  0.603 ms
		3  NODE-BE2 (z.z.z.z)  0.783 ms  0.755 ms  0.726 ms
		
		qkd@node-be1:~$ show crypto ikev2 sa
		Tunnel-id Local            Remote           fvrf/ivrf     Status 
		3         ROUTER-BE1 /500  ROUTER-BE2 /500  none/quantum  READY  
		Encr: AES-GCM, keysize: 256, PRF: SHA512,  Hash: None, DH Grp:20,
                                    Auth sign: PSK, Auth verify: PSK, QR
		Life/Active Time: 600/22 sec
		CE id: 22897, Session-id: 770
		Local spi: E118034DE56D0C3C  Remote spi: 51D8693E89375E21
	\end{lstlisting}
	\caption{The first traceroute represents the situation when there is no active IPSEC tunnel between NODE-BE1 and NODE-BE2. NODE-BE1 forwards the traffic through the DMVPN hub, NODE-BW1. Once the IPSEC tunnel between NODE-BE1 and NODE-BE2 has been set up, the traffic  will flow directly between both nodes, hence needing one hop less and reducing end-to-end latency in the second traceroute, which can be a relevant save in long distance connections.
		The router shows details on the IKEv2 SA, including its cryptographic parameters (AES-GCM 256 bits, DH group 20, authentication via preshared key and Quantum Resistance (QR) through the negotiation of PPKs).}
	\label{fig:Traceroute}
\end{figure*}

A crucial element of this architecture is the SDN control plane, which manages all cryptographic key delivery. The SDN controller is responsible for dynamically managing key material, from QKD devices or generated via PQC algoritms, and securely distributing it to the network endpoints. This approach supports end-to-end (E2E) key management across all five nodes and enables real-time reconfiguration of tunnel parameters based on traffic demands, as shown in Figure~\ref{fig:Traceroute}. The SDN layer abstracts cryptographic heterogeneity and provides a unified management interface for both QKD and PQC-based key material, enabling seamless interoperability and streamlined operations.

\begin{figure}[!htb]
	\begin{minipage}{0.48\textwidth}
		\centering
		\includegraphics[width=.98\linewidth]{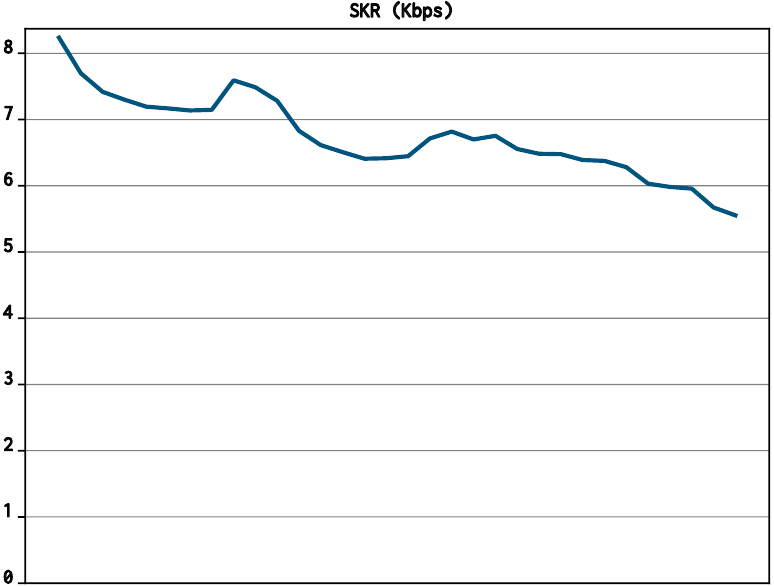}
		\caption{Key rate of LuxQuanta during all the field trial (Nova LQ).}
		\label{fig:luxquant}
	\end{minipage}\hfill
	\begin{minipage}{0.48\textwidth}
		\centering
		\includegraphics[width=.98\linewidth]{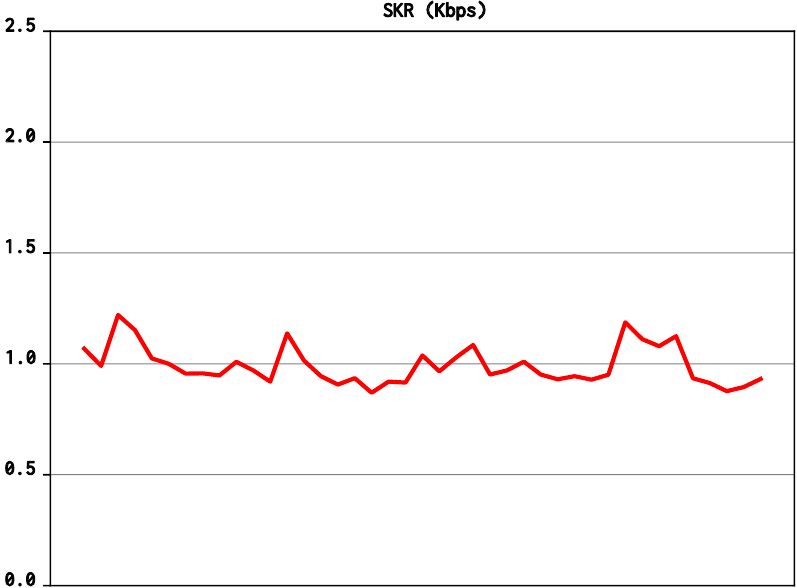}
		\caption{Key rate of IDQ link during all the field trial (Cerberis XG – 12dB).}
		\label{fig:idqkr}
	\end{minipage}
\end{figure}

The evaluation presented in this work is structured around three key groups of metrics, derived from real-world measurements taken across the five nodes. The first group of metrics are focused on all the quantum layer. Throughout the entire duration of the field trial, the QKD devices from both ID Quantique and LuxQuanta were continuously generating symmetric key material. These devices were integrated into the network as trusted key generation endpoints. Figures \ref{fig:luxquant} and \ref{fig:idqkr} illustrates the key generation rates of both vendors over a month and a half of the full field trial, demonstrating sustained performance and interoperability under realistic commercial network conditions.

The second group is focused on the throughput performance of IPsec tunnels, specifically the data transfer speeds achieved between every possible pair of Cisco routers configured in the network using PPKs. Each pair was tested under three different cryptographic modes: isolated classical Elliptic Curve Diffie–Hellman (ECDH) and hybridized ECDH + Quantum Key Distribution (HQKD) and ECDH + Post-Quantum Cryptography (HPQC), all methods applied directly on Cisco Catalyst 8300(V), with all tests executed using IPerf3 to ensure consistent and application-representative benchmarking. While throughput is not a metric directly related with the chosen key delivery method, it demonstrated the IPsec tunnels were operative during the field trial. These measurements reflect the actual bandwidth available to applications over fully secured tunnels.
\ref{figqkdiperf} and \ref{figsksperf}.

\afterpage{%
	\begin{figure}[p]
		\centering
		\includegraphics[width=1\linewidth]{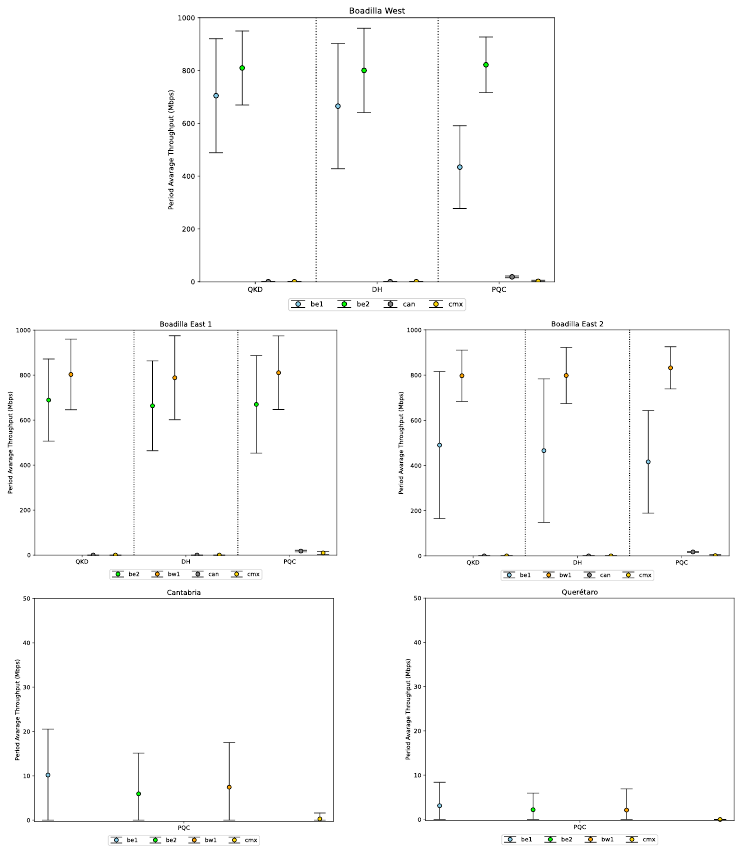}
		\caption{This figure summarizes the bandwidth performance of the quantum-safe network, based on IPerf measurements between key node pairs across four periods and their associated variance: HQKD-only, classical ECDH and cloud-connected HPQC nodes. At the top, the graph shows throughput from the central hub Boadilla West 1 in Madrid to the other four nodes, including the HPQC cloud endpoints in Cantabria and Querétaro. The middle section presents IPerf results from Boadilla East 1 (left) and Boadilla East 2 (right), while the lower section displays results from Cantabria (left) and Querétaro (right).  These detailed views allow for performance correlation across different paths and reveal the relative consistency and stability of QKD and PQC links. This comparison highlights the performance disparity between the high-speed, fiber-connected quantum trust domain and the more latency-constrained HPQC tunnels traversing long-haul networks and cloud ingress and egress firewall and related infrastructures, but both types of link, Quantum Safe protected. Together, the data confirms that, while cloud connections naturally exhibit lower throughput, secure communication remains sustained and reliable across all scenarios.}
		\label{figqkdiperf}
	\end{figure}
	\begin{figure}[p]
		\advance\leftskip-1.5cm
		\includegraphics[scale=1.2]{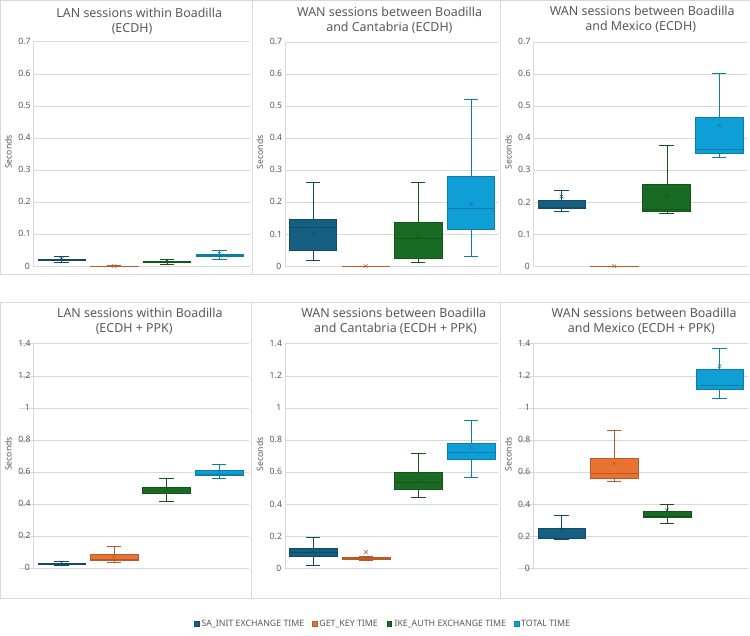}
		\caption{
			The latency impact of introducing quantum-safe mechanisms through our framework was evaluated by analysing 12,439 IKEv2 initial Security Association (SA) setups. Of these, 54\% relied solely on classical ECDH and were used as the baseline. The remaining sessions employed various combinations of QKD and PQC to generate RFC 8784 PPKs, enabling a full set of traffic samples.\\
			Introducing the KMS increased session-establishment latency by approximately 3× between Boadilla and Mexico (400 ms vs. 1,200 ms, with a 175 ms round-trip time (RTT)), 4× between Boadilla and Cantabria (200 ms vs. 800 ms, 7.5 ms RTT), and 15× within Boadilla itself (40 ms vs. 600 ms, 0.7 ms RTT). Most of this overhead arises from local get\_capabilities calls and from the interplay between the initiator and responder during get\_key operations. Specifically, the responder’s SKIP adapter benefits from the time consumed by the initiator’s GET\_KEY phase, partially compensating for latency effects.\\
			In absolute terms, the time required to establish new quantum-resistant SAs remains well within typical timeout thresholds for wide-area networks operating under real-world conditions, including jitter and packet loss.
		}
		\label{figsksperf}
	\end{figure}
}

The performance evaluation was conducted over a continuous period of two months. During this time, the network remained fully operational, with IPsec tunnels dynamically established using key material from the QKD systems, PQC algorithms and classic ECDH. The setup was only briefly interrupted to adjust different tests and get specific SD-QKD Nodes internal metrics as key material generation, key refresh rates, interoperability interface times, network key performance indicators like throughput or latencies, or to change the operational modes of the different nodes.

The results summarized in Figure \ref{figqkdiperf} present the throughput measurements obtained using IPerf3 across three distinct testing periods. The first period evaluates performance using HQKD keying exclusively, while the second assesses tunnels established through only classical ECDH exchange. Finally, the third period focuses entirely on HPQC-based communication. These periods offer a complete view of performance for quantum safe compared to classical settings. 

As it is shown in Figure \ref{figqkdiperf}, all nodes achieved throughput performances dependent on the bandwidth available, and latency between them. When in local area networks, the traffic was close to the capacity of the 1 Gpbs interfaces. With distant nodes the throughput is heavily impacted by latency and hardware limitations of the cloud-based test infrastructure. This throughput intervals were maintained during all the field trial, receiving fresh keys from the SDN layer every ten minutes. As indicated before, the key agreement scheme does not have a significant impact on the overall transmission throughput since most data traffic is encrypted with a symmetric algorithm, AES256 in this case. 

The HPQC segments spans all five nodes in the network, including the three Madrid-based locations interconnected via optical fiber and the two remote nodes. PQC-based key material was generated using ML-KEM. Secure communication between nodes was maintained using HPQC-protected key exchange.

The evaluation of the HPQC mode revealed performance levels closely aligned with those observed in the HQKD-secured segment of the network and classical ECDH exchange as it was expected, and it is shown on Figure \ref{figqkdiperf}. This similarity is largely attributed to the underlying SDN layer, which plays a central role in abstracting the key material nature from the key agreement mechanisms. Thanks to the integration of ML-KEM for key encapsulation within the provisioning mechanism to deliver it to the routers via the SKIP protocol, that is achieved with minimal overhead and not affecting the performance of the encryption layer. 

The measurements confirmed that the Quantum Safe key exchange method chosen to setup the IPsec tunnel does not affect performance. As a result, E2E IPsec tunnel setup was nearly indistinguishable from those recorded on the QKD/PQC of hybrid configurations. Moreover, thanks to the pre-provisioning of key material, rekeying operations proceed without delaying the setup, maintaining optimal system performance during key rotation events. This outcome reinforces the architectural principle that a well-designed key management layer, decoupled from the underlying cryptographic primitive (QKD, PQC or hybrid), can deliver consistent performance while supporting multiple quantum-safe technologies.

The third group of metrics analyzed corresponds to the Total IKEv2 SA Setup Time, evaluated as the sum of three intervals:
\begin{enumerate}
	\item SA\_INIT Exchange Time: Defined as the time between the generation of the SA\_INIT Request message and the completion of the SA\_INIT exchange at the Initiator.
	\item GET\_KEY Time: Defined as the time between the completion of the SA\_INIT exchange and the generation of the SA\_AUTH Request message at the Initiator. In ECDH-only sessions, without PPKs, this time is negligible and corresponds to fast internal processing within the router. When PPKs are negotiated, most of this time corresponds to the SKIP get\_key request.
	\item IKE\_AUTH Exhange Time: Defined as the time between the generation of the SA\_AUTH Request message and the creation of the SA at the Initiator router.
\end{enumerate}

As seen in Figure \ref{fig:skipsteps}a, the real total time for quantum-safe sessions would need to be incremented by the Initiator's get\_capabilities call, however the first log message including SA identifiers happens afterwards, when the SA\_INIT message is about to be generated. The get\_capabilites request is a local HTTPS call that typically lasts few tens of milliseconds. It could be approximated as slightly under the SA\_INIT Exchange Time in a LAN environment. Due to its minimal impact on the analysis, we preferred to stick to accurate measurements and avoid including this estimation.

This analysis allows a latency comparison between ECDH-only, where only the router to router messages in Figure \ref{fig:skipsteps}a exist, and ECDH + PPK sessions, which includes the calls to the SKIP interfaces at the Initiator and Responder ends plus all the internal processing in the KMS.

Figure \ref{figsksperf} shows the comparative analysis of the latencies in the different phases of the IKEv2 SA creation.

The upper row shows the times measured for the ECDH-only sessions, where the keys are negotiated without intervention of both the SKIP interface and the SDN control layer. The time to setup the IKEv2 SA was quite fast. Sessions using Cisco’s native IKEv2 implementation of ECDH average setup times around 40 milliseconds in Local Area Networks and up to 450 milliseconds between Madrid and Mexico, where we experienced an average round trip time of 175 milliseconds.

The lower row shows the times measured for the ECDH + PPK sessions, where the keys are negotiated with intervention of both the SKIP interface and the SDN control layer. The time to setup the IKEv2 SA was impacted by a noticeable latency. The total time needed to set up the IKEv2 SAs averaged 600 milliseconds in Local Area Networks and up to 1250 milliseconds between Madrid and Mexico.

In conclusion, sessions secured with PPKs have a noticeable increment in their setup latency that in any case stays well within the tolerances in real wide-area networks operating under real-world conditions.

It is important to remind that all the metrics related with the key delivery and rekeying operations were executed on a real production environment sharing resources with real production-grade traffic on both Local Area Network and Wide Area Network segments. This scenario represents a highly demanding environment, not only for the transport layer but also for the SDN-based key orchestration system. Despite these challenging conditions, the system consistently maintained service operation and successfully delivered E2E keys, enabling quantum-safe communications with no differential impact on client applications versus a ECDH-only setup.

In summary, the field deployment and subsequent testing validate the technical feasibility and operational efficiency of hybrid QKD and PQC architecture in a realistic banking communications network. The results confirm that our proposed setup can deliver quantum-resilient security without compromising performance or manageability. Furthermore, the use of SDN and SKIP provides a scalable and policy-driven platform for ongoing cryptographic evolution, making this architecture a viable candidate for future integration not only for banking but also for other critical infrastructures.

\section{Discussion}

Although this work is motivated by the requirements of the banking and financial sector, the architecture itself is domain‑agnostic. We use finance as a representative case because it exemplifies a highly regulated, security‑critical environment where operational constraints are unusually strict: rigorous compliance regimes, hard access‑control boundaries, and tightly governed physical and logical management of points of presence. In such settings, facility access is limited to personnel under formal procedures, and configuration changes are pre‑approved and always schedule, leaving little room for ad‑hoc intervention. The objective of the proof of concept is not to claim full conformance with a single end-to-end QKDN reference model, but to demonstrate a deployable integration path for banking environments. These realities favor non‑intrusive integration strategies in which new security capabilities—such as QKD and SDN needs to be introduced without disrupting existing infrastructure or requiring frequent manual reconfiguration, still, under this strict requirements, the experimental validation of our quantum-safe communication architecture demonstrates that it is technologically feasible to establish IPsec tunnels protected against future threats in banking infrastructures. The combination of different cryptographic key agreement schemes, vendors and interfaces is scalable in networks of critical sectors such as banking. The network was specifically designed with production requirements in mind, including high throughput, automation, cryptographic resilience, and seamless interoperability. By integrating the following seven main contributions, this work enables a structured roadmap for building next-generation, secure infrastructures that are prepared to resist both classical and quantum threats in production environments.


First, the network-oriented setup ensures that all user and application traffic is transparently encrypted through IPsec VPN tunnels. This architectural choice isolates the cryptographic complexity from the business logic, thereby preserving operational workflows and minimizing the cost of transition to quantum-safe technologies. By maintaining the encryption layer strictly at the VPN gateway level and relying on dynamic provisioning via SKIP and SDN, we eliminate the need for modification of client devices, core applications, or routing behaviors. This abstraction is critical in large-scale enterprise networks where heterogeneity and legacy systems are the norm.

Second, the scalability enabled by DMVPN, in combination with SDN-driven Quantum-safe key management, enables full-mesh tunnel dynamic IPsec tunnel creation in large networks, serving  on-time application connectivity demands, optimizing resources and easing the integration with the network’s security policy. This allows for ad-hoc, secure data exchanges between any pair of nodes, even across trust domains or geographic boundaries.


The third major contribution is the dynamic end-to-end quantum-safety architecture. Every tunnel established in the network is protected by at least one quantum-safe mechanism—either Post-Quantum Cryptography (PQC), Quantum Key Distribution (QKD), or both, in addition to ECDH. This diversity enables the system to defend against HNDL threats. The flexibility of combining QKD with PQC within the same control and delivery framework further enhances confidence in the cryptographic integrity of the communication adding a Defense in Depth layer.

Fourth, the architecture's multivendor interoperability addresses one of the most pressing challenges in adopting emerging cryptographic technologies: vendor lock-in. The experiment successfully integrated QKD systems from two independent vendors, and demonstrated seamless interoperability via ETSI compliant interfaces. This vendor agnostic management, facilitated by the SDN layer, proves that secure communication frameworks can evolve incrementally without dependency on proprietary standards, thus supporting market competitiveness and long-term adaptability.

Fifth, the network supports a multitechnology QKD deployment. The combination of DV-QKD and CV-QKD technologies provides architectural flexibility and future-proofing. DV-QKD, suitable for long-haul links with lower noise tolerance, and CV-QKD, optimal for metro-area high-rate deployments using standard optical components, were integrated into the same control plane. This capability allows network designers to select the most suitable QKD technology for each link based on distance, bandwidth, cost, and environmental conditions—without having to alter the overarching security model.

The sixth contribution is strong cryptographic agility, as it allows for rapid transitions between different security configurations, for example, in response to unforeseen events or operational requirements. Thanks to its SDN-based orchestration and standards-compliant interfaces, QKD, PQC, or hybrid schemes can be dynamically enabled, disabled, or reconfigured maintaining the services.

Finally, a strong Defense in Depth approach. The ability to dynamically combine cryptographic primitives—such as PQC and QKD on the same links, ensures that when possible, no tunnel relies solely on a single cryptosystem. This approach not only enhances resilience against cryptanalytic breakthroughs but also allows the system to adapt to evolving threat landscapes and standardization developments. As new quantum-safe algorithms become available or new vulnerabilities are discovered, the SDN controller can reconfigure keying mechanisms in real-time without interrupting service.

Overall, the experimental results validate that it is possible to build and operate a hybrid quantum-safe communication architecture with strong cryptographic assurances, high throughput, automated management, and vendor-neutral components. This work contributes not only a practical prototype but also a blueprint for broader industry adoption focused on financial infrastructures. Future extensions of this system could include QKD over satellite links or the integration with protocols beyond QKD like secure multiparty computation.

Several improvements have been identified to enhance the quality, observability and integration of the solution with existing operational frameworks. One key area is the incorporation of the Key Management Service and the QKD components into standard network-monitoring environments. Enabling the measurement of their performance parameters via SNMP would allow these systems to be supervised alongside conventional network devices, facilitating unified analysis and troubleshooting. Additionally, the logging capabilities of all solutions and devices can be enhanced. Introducing consistent contextual information, such as session identifiers or protocol state labels, across all relevant log messages would significantly improve traceability. It is also important to avoid formatting issues, such as unintended line breaks, that separate critical data from its associated context. Together, these refinements would provide clearer visibility into system behaviour and simplify post-event analysis.

This setup relied on the RFC 8784\cite{rfc8784} capabilities available in Cisco routers. As IPsec standards continue to evolve to incorporate post-quantum key-agreement mechanisms (such as RFC 9242\cite{rfc9242}, RFC 9370\cite{rfc9370}, draft-ietf-ipsecme-ikev2-mlkem\cite{ietf-ipsecme-ikev2-mlkem-03}, and, in particular, RFC 9867\cite{rfc9867}) we anticipate that future software releases for IPsec endpoints will include native support for ML-KEM and offer new options for incorporating external key material. Even in that scenario, the architecture presented here remains valuable for providing defense-in-depth in the key-agreement process by enabling hybridisation with external keys based on different security assumptions, including QKD or symmetric key-distribution systems. It may also serve as a practical vehicle for crypto-agility, allowing the integration of keys derived from alternative post-quantum key-agreement schemes, such as HQC\cite{HQC}, prior to their formal support in endpoint implementations.

As the transition to post-quantum security accelerates, architectures like the one presented here provide a viable path forward. They bridge the gap between theoretical cryptographic advances and operational deployments, enabling critical sectors like banking to protect their infrastructure and customer data in a rapidly evolving threat environment.

\ack{}

The authors wants to thank you the effort of UPM members Guillermo Vélez and Ramón Querol for the support and tech advise, and of the Santander team that collaborated to deploy and run the test scenario. Among them, Juan Rodríguez García, Pablo Zamorano Navarro, Roberto Cantalapiedra Escolar, Alberto Muñoz López, Carmen Yagüe Yagüe, Edgar Cañete Fernández, Javier García Asnedo and Carlos Galey Cózar, among others.

Also the gratefully acknowledge the support and collaboration of Cisco’s Country Digital Acceleration (CDA) Program, a global initiative that partners with governments and private sectors to drive sustainable, secure, and inclusive digital transformation. The CDA program’s innovative technology solutions and commitment to empowering communities have been instrumental in advancing the objectives of this project.

\funding{}

The authors would like to thank projects MADQuantum-CM, funded by Comunidad de Madrid (Programa de Acciones Complementarias) and by the Recovery, Transformation and Resilience Plan-Funded by the European Union-(NextGenerationEU, PRTR-C17.I1); the EU Horizon Europe project "Quantum Security Networks Partnership" (QSNP), grant 101114043; and EuroQCI-Spain, DEP grant 101091638; the QUBIP, funded by the European Union under the Horizon Europe framework program under grant agreement number 101119746, the PQ-REACT European Union’s Horizon Europe research and innovation programme under grant agreement number 101119547, European Union’s Digital Europe Programme under the project QUARTER (101091588), and from the European Innovation Council's Horizon Europe EIC Accelerator Programme under the project MIQRO (101161539).

\roles{}

R.J.V. formal analysis and experiment design, investigation, methodology, writing the original draft and technical developments. J.G.G. contributed to conceptualization, formal analysis and experiment design, investigation, methodology, supervision, validation, and writing the original draft. J.P.B. formal analysis and experiment design, investigation, methodology, supervision, validation and writing the original draft. Y.L. experiment design, methodology and technical developments. J.S.B. experiment design, methodology and technical developments. D.G. analysis of the results. M.A.S.S. methodology and project coordination. S.O. formal analysis and experiment design, methodology, validation and writing the original draft. S.G. experiment design, methodology and technical developments. J.S.P. formal analysis and experiment design, methodology, validation and writing the original draft. M.C formal analysis and experiment design, methodology, validation and writing the original draft. V.M. formal analysis and experiment design, investigation, methodology, supervision, validation and writing the original draft. All authors read the paper and contributed to the general readability of the whole paper. All authors read and approved the final manuscript.

\data{}

Not applicable.

\bibliographystyle{iopart-num-long}
\bibliography{bibliography}

@techreport{globalrisk,
	author = {Mosca, M. and Piani, M.},
	title = {Quantum Threat Timeline Report},
	institution = {Global Risk Institute},
	url = {https://globalriskinstitute.org/publication/2024-quantum-threat-timeline-report
	},
	urldate = {2025-07-31},
	year = {2025}
}

@techreport{FIPS203,
	author = {{National Institute of Standards and Technology}},
	title = {{Module-Lattice-Based Key-Encapsulation Mechanism Standard}},
	institution = {U.S. Department of Commerce},
	address= {Washington, D.C.},
	number = {Federal Information Processing Standards Publications (FIPS PUBS) 203},
	DOI = {10.6028/NIST.FIPS.203},
	year = {2024},
	month={August}
}

@techreport{FIPS204,
	author = {{National Institute of Standards and Technology}},
	title = {{Module-Lattice-Based Digital Signature Standard}},
	institution = {U.S. Department of Commerce},
	address= {Washington, D.C.},
	number = {Federal Information Processing Standards Publications (FIPS PUBS) 204},
	DOI = {10.6028/NIST.FIPS.204},
	year = {2024},
	month={August}
}

@ARTICLE{DH,
	author={Diffie, W. and Hellman, M.},
	journal={IEEE Transactions on Information Theory}, 
	title={{New directions in cryptography}}, 
	year={1976},
	volume={22},
	number={6},
	pages={644-654},
	doi={10.1109/TIT.1976.1055638}
}

@article{shor,
	author = {Shor, Peter W.},
	title = {{Polynomial-Time Algorithms for Prime Factorization and Discrete Logarithms on a Quantum Computer}},
	year = {1997},
	month = {October},
	issue_date = {Oct. 1997},
	publisher = {Society for Industrial and Applied Mathematics},
	address = {USA},
	volume = {26},
	number = {5},
	issn = {0097-5397},
	url = {https://doi.org/10.1137/S0097539795293172},
	doi = {10.1137/S0097539795293172},
	journal = {SIAM J. Comput.},
	pages = {1484–1509},
	numpages = {26}
}

@article{RSA,
	author = {Rivest, Ronald and Shamir, Adi and Adleman, Leonard},
	year = {1978},
	month = {01},
	pages = {120-126},
	title = {A Method for Obtaining Digital Signatures and Public-Key Cryptosystems},
	volume = {21},
	journal = {Commun. ACM},
	doi = {10.1145/357980.358017}
}

@inproceedings{MillerECC,
	author = {Miller, Victor},
	year = {1985},
	month = {01},
	pages = {417-426},
	title = {Use of Elliptic Curves in Cryptography.},
	booktitle = {Conference: Advances in Cryptology - CRYPTO '85, Santa Barbara, California, USA, August 18-22, 1985, Proceedings}
}

@article{peev2009secoqc,
	title={The SECOQC quantum key distribution network in Vienna},
	author={Peev, Momtchil and Pacher, Christoph and All{\'e}aume, Romain and Barreiro, Claudio and Bouda, Jan and Boxleitner, Winfried and Debuisschert, Thierry and Diamanti, Eleni and Dianati, Mehrdad and Dynes, James F and others},
	journal={New journal of physics},
	volume={11},
	number={7},
	pages={075001},
	year={2009},
	publisher={IOP Publishing}
}

@misc{oqs2025liboqs,
	title  = {liboqs - an open source C library for quantum-safe cryptographic algorithms},
	author = {{The Open Quantum Safe Project}},
	note = {{\url{https://github.com/open-quantum-safe/liboqs}}},
	url    = {https://github.com/open-quantum-safe/liboqs},
	year   = {2025}
}

@techreport{FIPS205,
	author = {{National Institute of Standards and Technology}},
	title = {{Stateless Hash-Based Digital Signature Standard}},
	institution = {U.S. Department of Commerce},
	address= {Washington, D.C.},
	number = {Federal Information Processing Standards Publications (FIPS PUBS) 205},
	DOI = {10.6028/NIST.FIPS.205},
	year = {2024},
	month={August}
}

@misc{Implementation_Roadmap,
  author       = {{European Commission}},
  title        = {A Coordinated Implementation Roadmap for the Transition to Post-Quantum Cryptography},
  howpublished = {\url{https://digital-strategy.ec.europa.eu/en/library/coordinated-implementation-roadmap-transition-post-quantum-cryptography}},
  year         = {2025},
  note         = {Accessed: 2025-11-17}
}

@misc{Timeline_Post_Quantum,
  author       = {{FS-ISAC}},
  title        = {The Timeline for Post Quantum Cryptographic Migration},
  howpublished = {\url{https://www.fsisac.com/the-timeline-for-post-quantum-cryptographic-migration}},
  year         = {2025},
  note         = {Accessed: 2025-11-17}
}

@misc{SDN_arch_ONF,
	author = {Betts, M. and Qiaogang, C. and Contreras-Murillo, L.M.and Davis, N. and Doolan, P. and Hood, D. and Janz, C. and Lam K. and Fengkai, L. and Paul, M. and Reith, L. and Schaller, S. and Schneider, F. and Shew, S. and Varma, E. and Vissers, M.},
	note = {\url{http://www.opennetworking.org}},
	publisher = {Open Networking Foundation},
	title = {{Publication ONF TR-521: SDN Architecture for Transport Networks 1.1}},
	year = {2016}
}

@ARTICLE{Kreutz_SDN_Survey,
	author={Kreutz, Diego and Ramos, Fernando M. V. and Veríssimo, Paulo Esteves and Rothenberg, Christian Esteve and Azodolmolky, Siamak and Uhlig, Steve},
	journal={Proceedings of the IEEE}, 
	title={{Software-Defined Networking: A Comprehensive Survey}},
	year={2015},
	volume={103},
	number={1},
	pages={14-76},
	doi={10.1109/JPROC.2014.2371999}}

@ARTICLE{Engineering_SDN_QKD,
	author={Aguado, Alejandro and Lopez, Victor and Lopez, Diego and Peev, Momtchil and Poppe, Andreas and Pastor, Antonio and Folgueira, Jesus and Martin, Vicente},
	journal={IEEE Communications Magazine}, 
	title={{The Engineering of Software-Defined Quantum Key Distribution Networks}}, 
	year={2019},
	volume={57},
	number={7},
	pages={20-26},
	doi={10.1109/MCOM.2019.1800763}
}

@Article{MadQCI,
	author={V. Martin and J. P. Brito and L. Ortiz and R. B. Mendez and J. S. Buruaga and R. J. Vicente and A. Sebastián-Lombraña and D. Rincon and F. Perez and C. Sanchez and M. Peev and H. H. Brunner and F. Fung and A. Poppe and F. Fröwis and A. J. Shields and R. I. Woodward and H. Griesser and S. Roehrich and F. De La Iglesia and C. Abellan and M. Hentschel and J. M. Rivas-Moscoso and A. Pastor and J. Folgueira and D. R. Lopez},
	title={{MadQCI: a heterogeneous and scalable SDN-QKD network deployed in production facilities}},
	journal={npj Quantum Information},
	year={2024},
	month={Sep},
	day={02},
	volume={10},
	number={1},
	pages={80},
	issn={2056-6387},
	doi={10.1038/s41534-024-00873-2},
	url={https://doi.org/10.1038/s41534-024-00873-2}
}

@techreport{ETSI_004,
	author = {ETSI},
	title = {{Quantum Key Distribution (QKD); Application Interface}},
	institution = {ETSI ISG},
	month = {August},
	number = {GS QKD 004 v2.1.1},
	year = 2020,
	url = {https://www.etsi.org/deliver/etsi_gs/QKD/001_099/004/02.01.01_60/gs_QKD004v020101p.pdf}
}

@techreport{ETSI_014,
	author = {ETSI},
	title = {{Quantum Key Distribution (QKD); Protocol and data format of REST-based key delivery API}},
	institution = {ETSI ISG},
	month = {February},
	number = {GS QKD 014 v1.1.1},
	year = 2019,
	url = {https://www.etsi.org/deliver/etsi_gs/QKD/001_099/014/01.01.01_60/gs_qkd014v010101p.pdf}
}

@techreport{ETSI-QSC-Report-2025,
	author = {ETSI},
	title = {{Preparing for a quantum secure future}},
	institution = {ETSI ISG},
	month = {April},
	year = 2025,
	url = {https://www.etsi.org/e-brochure/ETSI-QSC-Report-2025/mobile/index.html}
}

@techreport{ietf-ipsecme-ikev2-mlkem-03,
    number =    {draft-ietf-ipsecme-ikev2-mlkem-03},
    type =      {Internet-Draft},
    institution =   {Internet Engineering Task Force},
    publisher = {Internet Engineering Task Force},
    note =      {Work in Progress},
    url =       {https://datatracker.ietf.org/doc/draft-ietf-ipsecme-ikev2-mlkem/03/},
    author =    {Panos Kampanakis},
    title =     {{Post-quantum Hybrid Key Exchange with ML-KEM in the Internet Key Exchange Protocol Version 2 (IKEv2)}},
    pagetotal = 13,
    year =      2025,
    month =     sep,
    day =       29,
}

@techreport{rfc8784,
  author      = {Scott Fluhrer and Panos Kampanakis and David McGrew and Valery Smyslov},
  title       = {Mixing Preshared Keys in the Internet Key Exchange Protocol Version 2 (IKEv2) for Post-quantum Security},
  institution = {RFC Editor},
  series      = {Request for Comments},
  number      = {8784},
  howpublished= {RFC 8784},
  publisher   = {RFC Editor},
  doi         = {10.17487/RFC8784},
  url         = {https://www.rfc-editor.org/info/rfc8784},
  year        = {2020},
  month       = jun
}

@misc{rfc5996,
    series =    {Request for Comments},
    number =    5996,
    howpublished =  {RFC 5996},
    publisher = {RFC Editor},
    doi =       {10.17487/RFC5996},
    url =       {https://www.rfc-editor.org/info/rfc5996},
    author =    {Pasi Eronen and Yoav Nir and Paul E. Hoffman and Charlie Kaufman},
    title =     {{Internet Key Exchange Protocol Version 2 (IKEv2)}},
    pagetotal = 138,
    year =      2010,
    month =     sep,
}

@techreport{ETSI_015,
	author = {ETSI},
	title = {{Quantum Key Distribution (QKD); Control Interface for Software Defined Networks}},
	institution = {ETSI ISG},
	month = {April},
	number = {GS QKD 015 v2.1.1},
	year = 2022,
	url = {https://www.etsi.org/deliver/etsi_gs/QKD/001_099/015/02.01.01_60/gs_QKD015v020101p.pdf}
}

@article{leverrier_finite-size_2010,
	title = {Finite-size analysis of a continuous-variable quantum key distribution},
	volume = {81},
	author = {A. Leverrier, F. Grosshans, P. Grangier},
	journal = {Physical Review A},
	issn = {10502947},
	doi = {10.1103/PhysRevA.81.062343},
	month = jun,
	year = {2010},
	note = {arXiv: 1005.0339}
}

@article{Lim_Curty_2014,
	title = {Concise security bounds for practical decoy-state quantum key distribution},
	author = {Lim, Charles Ci Wen and Curty, Marcos and Walenta, Nino and Xu, Feihu and Zbinden, Hugo},
	journal = {Phys. Rev. A},
	volume = {89},
	issue = {2},
	pages = {022307},
	numpages = {7},
	year = {2014},
	month = {Feb},
	publisher = {American Physical Society},
	doi = {10.1103/PhysRevA.89.022307},
	url = {https://link.aps.org/doi/10.1103/PhysRevA.89.022307}
}

@article{IDQ_defenceindepth,
	title = {QKD: Part of a Defense-In-Depth Security Strategy},
	author = {Huttner, Bruno and Prisco John and Dukatz Carl and McGregor, Kirk and Wood Elizabeth},
	journal = {Phys. Rev. A},
	year = {2024},
	month = {May},
	publisher = {SRI International},
	url = {https://quantumconsortium.org/publication/qkd-part-of-a-defense-in-depth-security-strategy/}
}

@article{alia2024100,
	title={100 Gbps Quantum-safe IPsec VPN Tunnels over 46 km Deployed Fiber},
	author={Alia, Obada and Huang, Albert and Luo, Huan and Amer, Omar and Pistoia, Marco and Lim, Charles},
	journal={arXiv preprint arXiv:2405.04415},
	year={2024}
}

@article{pistoia2023paving,
	title={Paving the way toward 800 Gbps quantum-secured optical channel deployment in mission-critical environments},
	author={Pistoia, Marco and Amer, Omar and Behera, Monik R and Dolphin, Joseph A and Dynes, James F and John, Benny and Haigh, Paul A and Kawakura, Yasushi and Kramer, David H and Lyon, Jeffrey and others},
	journal={Quantum Science and Technology},
	volume={8},
	number={3},
	pages={035015},
	year={2023},
	publisher={IOP Publishing}
}

@misc{HSBC_1,
	author = {HSBC},
	title = {HSBC Quantum Protection for AI},
	url = {https://www.hsbc.com/news-and-views/news/media-releases/2023/hsbc-pioneers-quantum-protection-for-ai-powered-fx-trading},
	urldate = {2025-07-31},
	year = {2025}
}

@misc{HSBC_2,
	author = {HSBC},
	title = {HSBC Quantum Safe Tokenized Gold},
	url = {https://www.hsbc.com/news-and-views/news/media-releases/2024/hsbc-pilots-quantum-safe-technology-for-tokenised-gold},
	urldate = {2025-07-31},
	year = {2025}
}

@misc{SoftBank,
	author = {SoftBank Corp, Toshiba},
	title = {Field Experiment of IPsec QKD-VPN},
	url = {https://www.global.toshiba/ww/news/digitalsolution/2023/09/news-20230920-01.html
	},
	urldate = {2025-07-31},
	year = {2025}
}

@misc{singapour,
	author = {Singapour Goverment},
	title = {National Quantum Safe Network},
	url = {https://www.imda.gov.sg/about-imda/emerging-technologies-and-research/national-quantum-safe-network-plus
	},
	urldate = {2025-07-31},
	year = {2025}
}

@misc{luxemb,
	author = {Luxemburg QCI},
	title = {Lux4QCI},
	url = {https://lux4qci.eu/349-2/},
	urldate = {2025-07-31},
	year = {2025}
}

@misc{ncsc,
	author = {National Cyber Security Centre},
	title = {PQC migration timelines},
	url = {https://www.ncsc.gov.uk/guidance/pqc-migration-timelines},
	urldate = {2025-07-31},
	year = {2025}
}

@misc{ciscoppk,
	author = {Cisco},
	title = {Quantum Encryption PPK},
	url = {https://www.cisco.com/c/en/us/td/docs/routers/ios/config/17-x/sec-vpn/b-security-vpn/m-sec-cfg-quantum-encryption-ppk.html},
	urldate = {2025-07-31},
	year = {2025}
}

@article{Koblitz1987,
	acknowledgement = {#ack-nhfb#},
	author = {Koblitz, Neal},
	bibdate = {Tue Oct 13 08:06:19 MDT 1998},
	classcodes = {B0250 (Combinatorial mathematics); B6120B (Codes); C1160 (Combinatorial
	mathematics); C1260 (Information theory)},
	coden = {MCMPAF},
	corpsource = {Dept. of Math., Washington Univ., Seattle, WA, USA},
	fjournal = {Mathematics of Computation},
	issn = {0025-5718},
	journal = {Mathematics of Computation},
	month = jan,
	mrclass = {94A60 (11T71 11Y16 68P25)},
	mrnumber = {88b:94017},
	mrreviewer = {Harald Niederreiter},
	number = 177,
	pages = {203--209},
	referencedin = {Referenced in \cite[Ref. 3]{Wiener:1998:PCP}.},
	title = {Elliptic Curve Cryptosystems},
	treatment = {T Theoretical or Mathematical},
	volume = 48,
	year = 1987
}

@article{marksteiner2018resilience,
	title={On the Resilience of a QKD Key Synchronization Protocol for IPsec},
	author={{Stefan Marksteiner, Benjamin Rainer and Oliver Maurhart}},
	journal={arXiv preprint arXiv:1801.01710},
	year={2018}
}

@article{heydari2014integrating,
	title={Integrating IPsec within OpenFlow architecture for secure group communication},
	author={{Vahid Heydari Fami Tafreshi, Ebrahim Ghazisaeedi, et al}},
	journal={ZTE COMMUNICATIONS JOURNAL},
	volume={12},
	number={2},
	pages={41--49},
	year={2014},
	publisher={An International ICT R\&D Journal Sponsored by ZTE Corporation}
}

@inproceedings{li2015sdn,
	title={SDN-based access authentication and automatic configuration for IPsec},
	author={{Yunchun Li and Jutao Mao}},
	booktitle={2015 4th International Conference on Computer Science and Network Technology (ICCSNT)},
	volume={1},
	pages={996--999},
	year={2015},
	organization={IEEE}
}

@techreport{marin2021yang,
	title={A YANG data model for IPsec flow protection based on software-defined networking (SDN)},
	author={Marin-Lopez, R and Lopez-Millan, G and Pereniguez-Garcia, F},
	year={July 2008},
	institution={RFC 9061}
}

@techreport{bjorklund2016yang,
  author      = {Martin Bjorklund},
  title       = {The YANG 1.1 Data Modeling Language},
  institution = {Internet Engineering Task Force},
  year        = {2016}
}

@article{b.mendezQuantumResistantSoftware2026,
  title = {Quantum Resistant Software {{Defined-Networking IPsec}}, Enabling {{ITS}} Communication over {{IP}} Networks on Real Telco Infrastructures},
  author = {B. Mendez, Ruben and S. Buruaga, Jaime and P. Brito, Juan and Pastor, Antonio and R. Lopez, Diego and Martin, Vicente},
  year = 2026,
  month = may,
  journal = {Computer Networks},
  volume = {280},
  pages = {112171},
  issn = {1389-1286},
  doi = {10.1016/j.comnet.2026.112171},
  urldate = {2026-04-07}
}

@misc{rfc9242,
    series =    {Request for Comments},
    number =    9242,
    howpublished =  {RFC 9242},
    publisher = {RFC Editor},
    doi =       {10.17487/RFC9242},
    url =       {https://www.rfc-editor.org/info/rfc9242},
    author =    {Valery Smyslov},
    title =     {{Intermediate Exchange in the Internet Key Exchange Protocol Version 2 (IKEv2)}},
    pagetotal = 14,
    year =      2022,
    month =     may,
}

@misc{rfc9370,
    series =    {Request for Comments},
    number =    9370,
    howpublished =  {RFC 9370},
    publisher = {RFC Editor},
    doi =       {10.17487/RFC9370},
    url =       {https://www.rfc-editor.org/info/rfc9370},
    author =    {C. Tjhai and M. Tomlinson and G. Bartlett and Scott Fluhrer and Daniel Van Geest and Oscar Garcia-Morchon and Valery Smyslov},
    title =     {{Multiple Key Exchanges in the Internet Key Exchange Protocol Version 2 (IKEv2)}},
    pagetotal = 29,
    year =      2023,
    month =     may,
}

@misc{rfc9867,
    series =    {Request for Comments},
    number =    9867,
    howpublished =  {RFC 9867},
    publisher = {RFC Editor},
    doi =       {10.17487/RFC9867},
    url =       {https://www.rfc-editor.org/info/rfc9867},
    author =    {Valery Smyslov},
    title =     {{Mixing Preshared Keys in the IKE\_INTERMEDIATE and CREATE\_CHILD\_SA Exchanges of the Internet Key Exchange Protocol Version 2 (IKEv2) for Post-Quantum Security}},
    pagetotal = 12,
    year =      2025,
    month =     nov,
}

@misc{HQC,
  author       = {Aguilar Melchor, Carlos and Aragon, Nicolas and Bettaieb, Slim and Bidoux, Loic and Blazy, Olivier and Deneuville, Jean-Christophe and Gaborit, Philippe and Hauteville, Adrien and Zémor, Gilles},
  title        = {{HQC}: {Hamming Quasi-Cyclic}},
  year         = {2025},
  howpublished = {\url{https://pqc-hqc.org/}},
  note         = {Accessed: 2025-12-19}
}

@techreport{ITU-T_Y.QKD-IPSec-fr,
  author      = {{ITU-T}},
  title = {{Y.QKD-IPSec-fr} Framework for integration of quantum key distribution and IPsec},
  institution = {International Telecommunication Union},
  note = {{IETF} liaison statement to {ITU-T} {SG13} for progress on this work item},
  year = {2025},
  url = {https://www.ietf.org/lib/dt/documents/LIAISON/liaison-2025-03-24-itu-t-sg-13-opsawg-ls-on-work-progress-on-quantum-key-distribution-qkd-network-in-sg13-as-of-march-2025-attachment-32.pdf},
  publisher = {{IETF} | {Internet Engineering Task Force}},
}

@techreport{ITU-T_Y.3800,
  author      = {{ITU-T}},
  title       = {Recommendation {Y.3800}: {Overview on networks supporting quantum key distribution}},
  institution = {International Telecommunication Union},
  year        = {2019},
  month       = {oct},
  type        = {Recommendation},
  url         = {https://www.itu.int/rec/T-REC-Y.3800}
}

@techreport{ITU-T_Y.3801,
  author      = {{ITU-T}},
  title       = {Recommendation {Y.3801}: {Functional requirements for quantum key distribution networks}},
  institution = {International Telecommunication Union},
  year        = {2020},
  month       = {apr},
  type        = {Recommendation},
  url         = {https://www.itu.int/rec/T-REC-Y.3801}
}

@techreport{ITU-T_Y.3802,
  author      = {{ITU-T}},
  title       = {Recommendation {Y.3802}: {Quantum key distribution networks - Functional architecture}},
  institution = {International Telecommunication Union},
  year        = {2020},
  month       = {dec},
  type        = {Recommendation},
  url         = {https://www.itu.int/rec/T-REC-Y.3802}
}

@techreport{ITU-T_Y.3803,
  author      = {{ITU-T}},
  title       = {Recommendation {Y.3803}: {Quantum key distribution networks - Key management}},
  institution = {International Telecommunication Union},
  year        = {2020},
  month       = {dec},
  type        = {Recommendation},
  url         = {https://www.itu.int/rec/T-REC-Y.3803}
}

@techreport{ITU-T_Y.3804,
  author      = {{ITU-T}},
  title       = {Recommendation {Y.3804}: {Quantum key distribution networks - Control and management}},
  institution = {International Telecommunication Union},
  year        = {2020},
  month       = {sep},
  type        = {Recommendation},
  url         = {https://www.itu.int/rec/T-REC-Y.3804}
}

@techreport{ITU-T_Y.3805,
  author      = {{ITU-T}},
  title       = {Recommendation {Y.3805}: {Quantum key distribution networks - Software-defined networking control}},
  institution = {International Telecommunication Union},
  year        = {2021},
  month       = {dec},
  type        = {Recommendation},
  url         = {https://www.itu.int/rec/T-REC-Y.3805}
}

@techreport{ITU-T_X.1712,
  author      = {{ITU-T}},
  title       = {Recommendation {X.1712}: {Security requirements and measures for quantum key distribution networks - key management}},
  institution = {International Telecommunication Union},
  year        = {2021},
  month       = {oct},
  type        = {Recommendation},
  url         = {https://www.itu.int/rec/T-REC-X.1712}
}

\end{document}